\documentclass[usenatbib]{mnras}

\usepackage{graphicx}

\title[Assembly of bulge-dominated galaxies]{The assembly of bulge-dominated galaxies: two evolutionary channels traced through morphology, kinematics, and environment with a hybrid classification pipeline}

\author[Flores-Freitas et al. 2026]{R. Flores-Freitas,$^{1}$\thanks{E-mail: rodrigoff.astro@gmail.com}
R. R. de Carvalho,$^{1,2}$
K. V. Nedkova,$^{3}$
I. Ferreras,$^{4}$
I. Kolesnikov,$^{1}$
\newauthor V. M. Sampaio$^{5,6}$
\\
$^{1}$ Núcleo de Astrofísica, Universidade Cidade de São Paulo, Rua Galvão Bueno 868, 01506-000 SP, Brazil\\
$^{2}$ Universidade de São Paulo, IAG, Rua do Matão 1226, Cidade Universitária, São Paulo 05508-900, Brazil\\
$^{3}$ IPAC, California Institute of Technology, 1200 E.~California Blvd, Pasadena, CA 91125, USA\\
$^{4}$ Instituto de Astrofísica de Canarias, Vía Láctea, 38205 La Laguna, Tenerife, Spain\\
$^{5}$ Instituto de Física, Universidad Técnica Federico Santa María, Av. España 1680, Valparaíso, Chile\\
$^{6}$ Millennium Nucleus for Galaxies (MINGAL), Chile\\
}

\date{Accepted 2026 September 16. Received 2026 August 20; in original form 2026 June 20}
\pubyear{2026}

\begin{document}
\maketitle

\begin{abstract}
We investigate the physical origin of the bimodality in bulge-dominated galaxies, originally identified by Sampaio et al. (2025), by combining non-parametric morphological metrics, structural scaling relations, stellar kinematics, and environmental trends across a wide redshift range ($0.2 < z < 2.4$). Using the MEGG-based hybrid classification pipeline applied to CANDELS imaging, we examine the distributions of morphological metrics for two families of bulge-dominated galaxies: \emph{G1}, with high specific star formation rate (sSFR) distributions, similar to discs, and \emph{G2} with lower sSFR. We find that \emph{G1} galaxies occupy an intermediate position between discs and \emph{G2} spheroids in morphological metrics, and this behaviour persists up to $z = 1.4$. Fitting the Kormendy relation separately for each family, we find a persistent offset in the zero-point across all redshifts: \emph{G2} galaxies are systematically brighter in mean effective surface brightness at fixed effective radius, likely indicating higher central stellar densities. This offset is present in both observed and rest-frame magnitudes, and we argue that it reflects a genuine difference in assembly history. A cross-match with MUSE observations reveals that \emph{G1} galaxies have higher projected angular momentum than \emph{G2} galaxies, with \emph{G1} galaxies overlapping with the disc population, while \emph{G2} galaxies are more dispersion-dominated. The redshift evolution of morphological fractions shows that, at high stellar masses, the \emph{G2} fraction grows, while discs follow the opposite trend. In parallel, \emph{G1} remains a stable, lower-mass population consistent with secular bulge growth. Finally, within galaxy clusters, \emph{G1} galaxies are preferentially found at larger cluster-centric radii, suggesting that high-density environments amplify the bimodality by accelerating quenching.
\end{abstract}


\section{Introduction}

The connection between galaxy morphology and star formation has long been understood through the classical dichotomy of disc, star-forming systems and spheroidal, passively evolving galaxies. In this framework, spheroids (whether bulges, elliptical galaxies, or compact systems) are traditionally associated with old stellar populations, high central stellar mass concentrations, and low gas fractions, reflecting a formation pathway dominated by violent relaxation, dissipative collapse, or mergers. However, over nearly three decades, a substantial body of observational and theoretical work has quietly revealed the existence of \emph{star-forming spheroids}, a population that challenges the neat separation between structural morphology and star-formation activity \citep[e.g.][]{Menanteau2001, Schawinski2009, Dekel2014}. These findings, originating from disparate surveys, redshift ranges, and methodological approaches, have remained fragmented in the literature, preventing the emergence of a unified physical picture. Only recently, with the analysis presented in \citet{Sampaio2025}, was bimodality in bulge-dominated galaxies across a wide redshift ($z$) range shown, revealing two distinct evolutionary families. Before discussing this development, it is essential to review the broad, scattered literature that has hinted at the existence, prevalence, and physical significance of star-forming spheroids over cosmic time.

The earliest indications that spheroids may undergo substantial star formation were found in high-redshift observations targeting the epoch when massive galaxies formed. Using \textit{HST} imaging of a protocluster at $z \approx 2.4$, \citet{Pascarelle1996} identified compact, actively star-forming systems which they interpreted as nascent spheroids. These objects exhibited the high surface brightness and compactness typical of local spheroids, yet with ultraviolet morphologies and star formation rates inconsistent with passive evolution. Their discovery provided the first observational evidence that spheroidal structure could arise concurrently with, rather than subsequent to, episodes of intense star formation.

Theoretical models developed shortly thereafter reinforced this interpretation. Semi-analytic galaxy formation models aiming to reproduce infrared and submillimetre source counts found it necessary to include a population of \emph{dusty, star-forming spheroids} undergoing rapid stellar mass assembly \citep{Perrotta2003,Silva2004,Silva2005}. In these models, spheroids formed most of their stars in short-lived yet extremely intense bursts, often associated with gas-rich mergers or collapse-driven events. While these models successfully reproduced several observational constraints, they did not explore in detail whether star-forming spheroids remained structurally distinct from disc galaxies or how they might be connected to the properties of quenched spheroids at later times. Nevertheless, the theoretical consensus emerged that active star formation within spheroidal systems was not only plausible but likely common during the early Universe.

A separate line of evidence emerged at intermediate and low redshifts, where deep imaging surveys allowed more refined morphological classifications. \citet{Im2001} identified a class of objects they termed ``blue spheroids''-galaxies with morphological classifications consistent with E/S0 types but exhibiting blue colours and young stellar populations. 
This evidence was extended to a wider redshift baseline using deep HST imaging of the GOODS fields, where a substantial fraction of morphologically classified early-type galaxies were found to display blue cores and ongoing or residual star formation across $0.4 < z < 1.5$, with the star-forming fraction rising toward lower stellar masses \citep{Ferreras2005,Ferreras2009a,Ferreras2009b}. This discovery challenged the long-standing assumption that spheroidal morphology unambiguously signals quiescence. 

Subsequent analyses of local early-type galaxies strengthened this picture. \citet{Helmboldt2008} and \citet{Shapiro2010} demonstrated that a significant fraction of morphologically classified early-type galaxies show evidence of residual star formation, cold gas reservoirs, or disc-like kinematic components. These findings suggested that early-type morphology encompasses a structurally diverse population whose star formation histories cannot be uniformly characterised by a single quenching event.

Further evidence accumulated with the identification of star-forming early-type or spheroidal systems in large surveys such as GAMA \citep{Driver2009}. \citet{George2015} and \citet{Mahajan2018} showed that blue spheroids occupy a wide range of masses, from dwarf to massive systems, and often share structural properties—Sérsic indices, effective radii, and concentration parameters—with classical spheroids. Yet they lie on the star-forming main sequence, indicating that they have not completed their stellar mass assembly. \citet{Moffett2019} analysed rotating spheroidal galaxies and concluded that many are dynamically and morphologically spheroidal while maintaining significant star formation, often supported by a residual disc or rotational component.

Large statistical studies expanded this interpretation. \citet{Paspaliaris2023} analysed the star formation properties of morphologically classified early-type galaxies and found that nearly half remain star-forming, a result that challenges the traditional view of the early-type population as uniformly passive. Meanwhile, \citet{Kim2018}, using SDSS imaging and spectroscopy, demonstrated a structural bifurcation: \emph{star-forming spheroids have systematically lower Sérsic indices than quenched spheroids}. This finding suggests that spheroids undergoing star formation may retain structural links to disc components or form through different processes than fully quenched spheroids.

Environmental studies provided yet another dimension of complexity. \citet{Kuchner2017} examined the stellar mass–size relation of cluster galaxies and identified a mixed population of star-forming spheroids whose sizes and masses differ from quenched spheroids. Their results indicated that spheroidal morphology can be established prior to quenching, supporting a scenario in which morphological transformation and star formation suppression need not occur simultaneously or through the same mechanism, in agreement with the results shown in \cite{Sampaio2024} and \cite{Sampaio2025}.

A parallel but complementary line of evidence for diversity within the spheroid population came from integral-field spectroscopy (IFS). The SAURON and ATLAS$^{\rm 3D}$ surveys \citep{Emsellem2007,Emsellem2011,Cappellari2011} demonstrated that early-type galaxies divide into two kinematically distinct classes - fast and slow rotators - with the slow-rotator fraction rising sharply in dense environments and at the highest masses, and with slow rotators tending to occupy the cores of groups and clusters. Subsequent wide-field IFS surveys, including SAMI \citep{vanDeSande2021} and MaNGA \citep{Graham2018}, extended this result to larger statistical samples, confirming that projected specific angular momentum is a more physically informative classifier than morphology alone. The kinematic morphology–density relation \citep{Cappellari2011,Deugenio2013} showed that environment does not merely change the number of spheroids, but selectively populates cluster cores with the most dynamically hot systems. These findings established that spheroidal morphology alone is insufficient to characterise the assembly history of a galaxy, and that angular momentum serves as an independent diagnostic of how structural transformation was achieved, a conclusion that is directly relevant to the bimodality explored in this work.

The advent of deep imaging from the Cosmic Assembly Near-infrared Deep Extragalactic Legacy Survey \citep[CANDELS;][]{Grogin2011,Koekemoer2011} and the even higher resolution of the James Webb Space Telescope (JWST) have further pushed the boundary of what can be identified as a spheroidal system. Several studies have shown that star-forming galaxies at $z > 2$ can fall within the spheroidal region of non-parametric morphology spaces such as Gini–$M_{20}$ \citep{Law2012,Lee2013}. This behaviour suggests that central concentration—a key indicator of spheroidal structure—may develop before or concurrently with the quenching of star formation. Findings of compact star-forming bulges or spheroids have been linked to ``compaction'' scenarios, in which galaxies undergo centrally concentrated starbursts before quenching and size growth. Although these compact, star-forming systems share qualitative similarities with the blue spheroids at lower redshift, their interpretation has remained focused on high-redshift assembly and quenching pathways rather than identifying a persistent, long-lived class of star-forming spheroids.

The physical mechanisms that can produce these distinct spheroidal families have been extensively modelled. Merger-driven scenarios, in which gas-rich major mergers drive dissipative collapse, efficiently destroy angular momentum, build compact remnants with high central stellar densities, trigger AGN feedback, and lead to rapid quenching - providing a natural pathway to the classical, dispersion-dominated spheroid \citep{Hopkins2006,Hopkins2008,Naab2014}. At the same time, the compaction scenario \citep{Dekel2014,Zolotov2015,Tacchella2016} proposes that wet compaction events (driven by counter-rotating streams, minor mergers, or violent disc instabilities) can produce compact, centrally star-forming objects that subsequently quench inside-out, leaving a dense stellar core. Both pathways predict a rapid rise in the central stellar mass surface density as a precursor to quenching \citep{Cheung2012,Fang2013,Barro2017}, making it a useful tracer of evolutionary state. By contrast, secular evolution (involving bar-driven inflows, minor perturbations, and slow disc instabilities) builds bulge mass while preserving angular momentum and sustaining star formation \citep{KormendyKennicutt2004,FisherDrory2016}. Furthermore, the formation of spheroidal galaxies does not necessarily begin with a disc structure, as scenarios have also been proposed in which the spheroid forms first, and a disc grows around it if specific conditions are later met \citep{Cook2009,Mo2024}. The coexistence of these channels implies that bulge-dominated galaxies at intermediate and high redshifts should not form a homogeneous population, and that structural diagnostics and kinematic tracers should reflect the imprint of the dominant assembly pathway. 

Despite these advances, a comprehensive framework that connects structural properties and star formation activity across the epoch of peak bulge assembly, from $z \approx 2.5$ to the present, without relying on visual classifications, was only recently presented in \citet{Sampaio2025}. Using bulge-dominated galaxies in the redshift range $0.2 < z < 2.4$, they identified a striking bimodality in specific star formation rates, stellar masses, and structural parameters among spheroids. This bimodality defines two distinct families: \emph{G1}, a long-lived population of less massive galaxies with lower Sérsic indices, extended light distributions, and ongoing or residual star formation; and \emph{G2}, a massive galaxy population with significantly less star formation and whose structural properties (high Sérsic indices, compact light profiles) suggest formation through merger-driven or dissipative processes. Unlike most previous studies, which probe a single redshift interval or a single structural parameter at a time, \citet{Sampaio2025} demonstrate that these two distinct families of bulge-dominated galaxies persist over several gigayears of cosmic history.

This unified perspective positions star-forming spheroids not as anomalies or transitional objects but as a fundamental component of bulge evolution. \emph{G1} spheroids connect naturally to the star-forming early types observed throughout the literature. \emph{G2} spheroids align with the classical quenched population, whose structural and dynamical properties are consistent with a merger-driven origin. The coexistence and divergent evolution of these two spheroidal families offer a physically motivated explanation for decades of observational findings and establish a new paradigm for understanding bulge formation and the morphological transformation of galaxies across cosmic time.

In this work, we build upon the bimodality identified by \cite{Sampaio2025} to investigate the physical origin and evolutionary significance of the two bulge-dominated galaxy families, \emph{G1} and \emph{G2}, across the redshift range $0.2 \leq z \leq 2.0$. Using an unsupervised-supervised hybrid classification pipeline grounded in the MEGG morphological system \citep{Kolesnikov2025} applied to CANDELS imaging, we characterise how the two families differ in non-parametric morphological metrics, in the zero-point of the Kormendy relation computed from both observed and rest-frame magnitudes, and in projected specific angular momentum derived from IFS cross-matches. We further combine our sample with a large cluster catalogue to quantify how the environment modulates the relative spatial distribution of \emph{G1} and \emph{G2} galaxies, and we track the redshift evolution of morphological fractions to constrain the assembly channels that connect the two populations to the broader galaxy population across cosmic time.

This paper is organised as follows: in Section \ref{sec:data_methods} we describe the galaxy sample, the morphological classification procedure, and the ancillary catalogues used to derive structural parameters, spin parameters, and cluster membership; in Section \ref{sec:results} we present the distributions of morphological metrics, the Kormendy relation offsets, the rotational support analysis, and the environmental trends for \emph{G1} and \emph{G2} galaxies; in Section \ref{sec:discussion} we discuss the evolutionary channels implied by these results and their connection to the classical/pseudo-bulge dichotomy; finally, in Section \ref{sec:conclusions} we summarise our main conclusions. Throughout this paper, we adopt a flat $\Lambda$CDM cosmology, with $\Omega_{\rm m,0} = 0.3089$, $\Omega_{\rm \Lambda,0} = 0.6911$, $\Omega_{\rm b,0} = 0.0486$, and $H_0 = 67.74$ km\,s$^{-1}$\, Mpc$^{-1}$, from \cite{Planck2016} and report magnitudes in the AB system.


\section{Data and Methods}
\label{sec:data_methods}

\subsection{Galaxy Sample}
\label{sec:galaxy_sample}
The main galaxy sample used in this work is from \cite{Sampaio2025} and comprises galaxies from CANDELS \citep{Grogin2011,Koekemoer2011}. These are selected from the COSMOS, UDS, EGS and GOODS-S fields, which have similar photometric characterisation and stellar mass estimation methods. To avoid unreliable morphological classifications of faint sources \citep{Grogin2011,Kartaltepe2015}, a magnitude cut is adopted: only galaxies brighter than 24 magnitudes in the \textit{H}-band (F160W) are selected. The magnitude selection criteria yield a sample of 30193 galaxies, which is further reduced to 26509 after excluding galaxies without observations in the F814W filter. 

The stellar masses for the selected galaxies are retrieved from \cite{Santini2015} and \cite{Barro2019}. To avoid objects with unreliable stellar mass ($M_\ast$) estimates, \cite{Sampaio2025} selects only galaxies with $\log (M_\ast/{\rm M_\odot}) \geq 9$, reducing the sample to 14776 galaxies. These galaxies are further divided into spheroids, discs and irregular galaxies using the morphological classification method described in Section \ref{sec:morph_classification}. As presented in \cite{Sampaio2025}, the spheroid population exhibits bimodality in sSFR over a wide redshift range and is therefore divided into two families of galaxies, called \emph{G1} and \emph{G2}. This division is made through a decomposition of the sSFR distribution of spheroids using individual Gaussian mixture models at different redshift bins. This decomposition reveals two distinct classes of bulge-dominated galaxies that have markedly different sSFR distributions. \emph{G1} galaxies have sSFR distributions that are consistent with disc galaxies, while \emph{G2} spheroids have significantly lower sSFRs. We emphasise here that only the sSFR distribution is used to distinguish \emph{G1} from \emph{G2} spheroids. In other words, these two families are not directly separated by morphology, but only by their star formation activity. We reinforce this point to highlight that the morphological classification is blind to sSFR in our framework, and that the \emph{G1}/\emph{G2} classes are identified solely by the bimodality of sSFR.

In this work, we consider only galaxies with robust morphological classifications and limit our analysis to galaxies at $0.2 \leq z \leq 2.4$. Thus, our CANDELS sample comprises 13049 galaxies (7430 discs, 2139 \emph{G1} spheroids, 2189 \emph{G2} spheroids, and 1291 irregulars). In Sections \ref{sec:KR_method}, \ref{sec:cluster_crossmatch} and \ref{sec:crossmatch_lambdaR} we describe how we combine this final sample with other catalogues available in the literature to generate subsamples. To facilitate understanding of the sizes of the different subsamples in this work, we present in Table \ref{tab:subsamples} the number of galaxies used in each analysis presented in Section \ref{sec:results}.

\begin{table*}
    \centering
    \begin{tabular}{c|c|c|c} \hline
        Analysis subject & Number of galaxies & Description & Redshift interval \\
             & (\emph{G1} / \emph{G2} / Disc) &             &         \\ \hline
        Morphological Metrics & 2139 / 2189 / 7430 & Section \ref{sec:galaxy_sample} & $0.2 \leq z \leq 2.4$ \\ 
        Kormendy Relation and Sérsic Parameters (F814W) & 1592 / 1310 / 4617 & Section \ref{sec:KR_method} & $0.2 \leq z \leq 2.4$ \\
        Kormendy Relation (J-band) & 400 / 317 / 933  & Section \ref{sec:KR_method} & $0.2 \leq z \leq 2.4$ \\
        Cluster Environment & 202 / 431 / 996  & Section \ref{sec:cluster_crossmatch} & $0.2 \leq z \leq 1.4 $ \\
        Spin Parameter & 11 / 18 / 17  & Section \ref{sec:crossmatch_lambdaR} & $0.2 \leq z \leq 0.8$ \\ \hline
    \end{tabular}
    \caption{Summary of galaxy samples analysed in this work. From left to right, the columns are: the subject of analysis for which the sample was used, the number of each type of galaxy in the sample, the section in which the sample is described, and the redshift interval in which the sample is analysed.}
    \label{tab:subsamples}
\end{table*}

\subsection{Morphological Classification}
\label{sec:morph_classification}

The morphological classification of galaxies is obtained through a hybrid unsupervised-supervised framework and is based on four non-parametric morphological metrics from the MEGG system \citep{Kolesnikov2025}. This system comprises the second moment of light \citep[$M_{20}$,][]{Lotz2004}, Shannon entropy \citep[$E$,][]{Bishop2007,Ferrari2015}, the Gini coefficient \citep{Abraham2003,Lotz2004}, and gradient pattern asymmetry \citep[$G_{2}$,][]{Rosa2018}, and the classification method uses the four-dimensional space of these metrics to efficiently separate disc and spheroidal galaxies without the biases of visual classification.
In what follows, we summarise the method, and for a comprehensive description of the pipeline, we refer the reader to \cite{Kolesnikov2024}, \cite{Kolesnikov2025} and \cite{Sampaio2025}. In brief, the hybrid pipeline leverages the intrinsic physical properties encapsulated within non-parametric morphological metrics. Based exclusively on the MEGG system, galaxies are initially pseudo-labelled via unsupervised clustering employing the Self-Organizing Maps (SOM) algorithm \citep{Kohonen1990} implemented in the \textsc{sombrero} package \citep{VillaVialaneix2017,Vialaneix2025}. In the subsequent phase, these pseudo-labelled samples are utilised to train a Convolutional Neural Network (CNN) ensemble, which in turn generates the final morphological classifications. The robustness and stability of this framework were rigorously tested and confirmed by \cite{Kolesnikov2024}, who applied the pipeline to a low-redshift, visually classified sample from Galaxy Zoo \citep{Lintott2011}, achieving an overall accuracy of approximately 90 per cent across more than half a million objects. Additional validation by \cite{Kolesnikov2025} demonstrated the method's stability against image-degradation effects introduced at high redshift, such as surface-brightness dimming and a decreasing angular scale. In essence, the inferred labels accurately represent intrinsic galactic structure, translating quantifiable structural metrics into robust morphological categories without relying on visual priors. Additionally, to prevent confounding effects from mixing galaxies across different cosmic epochs, the pipeline employs dedicated models for redshift bins spanning $0.2 \leq z \leq 2.4$ in steps of 0.2, as in \cite{Kolesnikov2025}. Each redshift bin is treated as a separate dataset, and the classification pipeline operates on it independently. This approach guarantees that the CNN models are trained and evaluated exclusively on galaxies occupying the same redshift interval. To reinforce the robustness of the morphological classification adopted here, in Appendix \ref{app:purity_contamination} we present estimates of the expected contamination of discs into spheroids and vice versa.

\begin{figure*}
    \centering
    \includegraphics[width=0.99\textwidth]{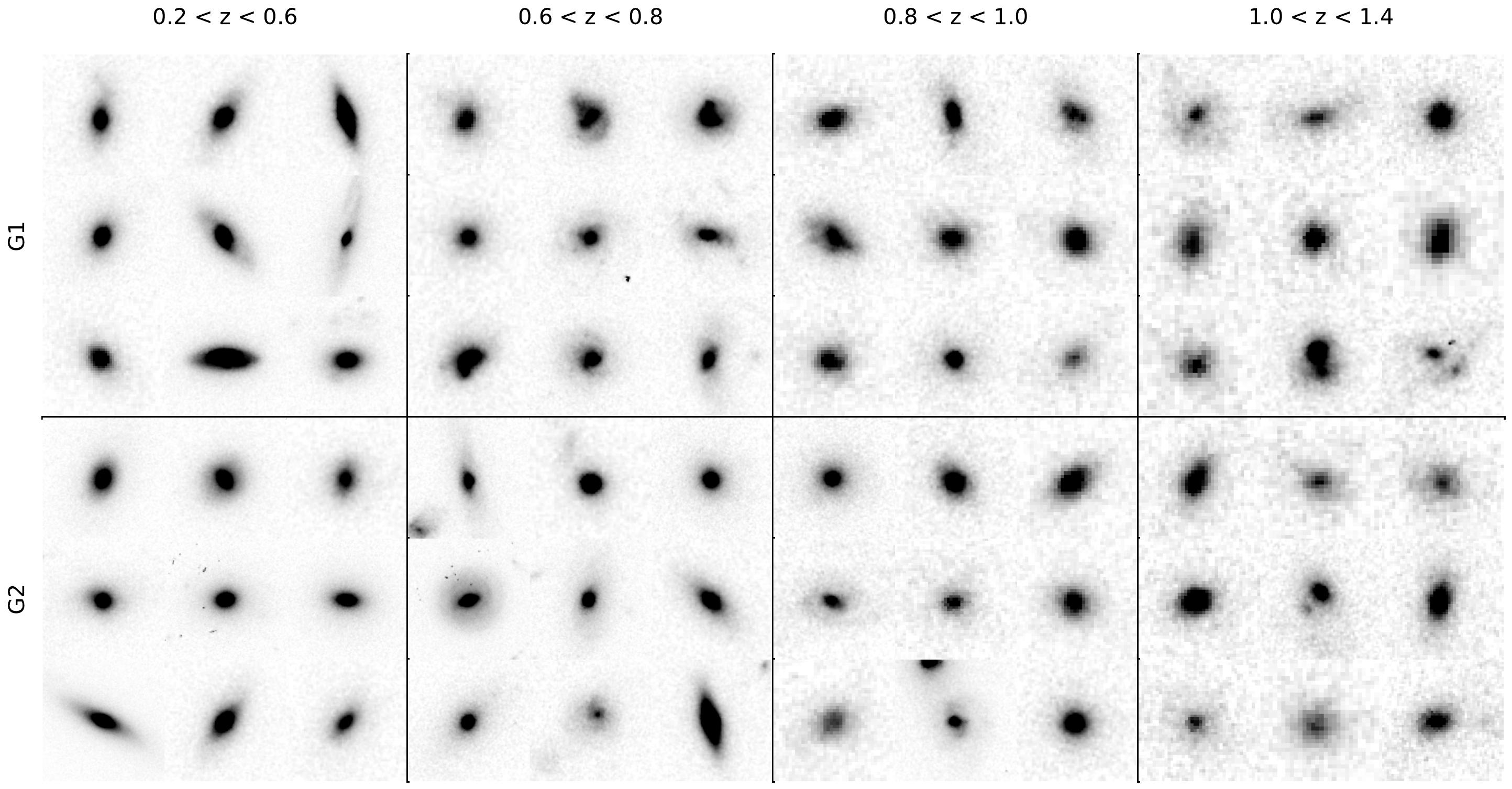}
    \caption{Examples of both classes of bulge-dominated galaxies (\emph{G1} and \emph{G2}) observed in the F814W filter in different redshift bins. These galaxies were randomly selected from the full dataset of \emph{G1} and \emph{G2} galaxies. Each square cutout is three Petrosian radii in size.}
    \label{fig:examples}
\end{figure*}

In this work, we focus on the bimodality of galaxies classified as spheroids in the aforementioned classification scheme, and in Figure \ref{fig:examples} we show examples of them. The spheroids, further classified as \emph{G1} and \emph{G2} (see Section \ref{sec:galaxy_sample}), are treated here as \textit{bulge-dominated} galaxies rather than \textit{pure} spheroids. This clarification of terminology is necessary to avoid confusion, since the term spheroid could lead the reader to infer that these galaxies have negligible rotational support or that we are referring only to the bulge substructure of galaxies. For the sake of clarity, it is important to note here that when referring to early-type galaxies, we are treating them as a larger class that \textit{contains} the bulge-dominated galaxies studied here.

\subsection{Kormendy relation, Sérsic parameters and rest-frame magnitudes}
\label{sec:KR_method}
To explore the \emph{G1} and \emph{G2} galaxy populations in the context of structural scaling relations, we fit the Kormendy relation \citep[KR; ][]{Kormendy1977} independently for each type of bulge-dominated galaxy in different redshift bins. The mean surface brightness magnitudes ($\langle \mu \rangle_e$) are computed in two forms, once using observed magnitudes in the F814W band and once using rest-frame magnitudes at the J-band (centred at 1.25 $\mu$m). The necessary magnitudes and half-light radii are obtained through cross-matches with the UVCANDELS \citep{Nedkova2024} and \cite{Santini2015} catalogues. Using \textsc{GALFITM} \citep{Haubler2013}, \cite{Nedkova2024} perform multi-wavelength single Sérsic two-dimensional profile fitting, utilising images from 7 to 10 of the filters available for CANDELS fields (F275W, F435W, F606W, F775W, F814W, F850LP, F105W, F125W, F140W, F160W). In their methodology, galaxy model parameters are replaced by wavelength-dependent functions \citep{Haubler2013,Haubler2022}, more specifically, Chebyshev polynomials of the first kind \citep{Abramowitz1965}. Thus, with the polynomial coefficients in their catalogue, we retrieve the values of total magnitudes, Sérsic indices and half-light radii at specific observed wavelengths. 

Our crossmatch with the \cite{Nedkova2024} catalogue is performed using right ascension (RA) and declination (DEC) coordinates, with a maximum separation of 0.5 arcsec, resulting in 10247 galaxies at $0.2 \leq z \leq 2.4$. Next, we remove galaxies from the sample based on the results of the Sérsic fits in the F814W filter, more specifically: galaxies where the profile fitting fails\footnote{The cases where \texttt{FLAG\_GALFIT!=2}, which means that either there are not enough bands for the multi-wavelength fitting or \textsc{GALFITM} did not converge.}, galaxies where the recovered radius is smaller than 2 pixels, and galaxies where the Sérsic index ($n$) is close to the constraints used in the fitting, that is, $n<0.25$ or $n>11.5$. Furthermore, we also remove galaxies with unphysical sizes, that is, those with a half-light radius greater than 30 kpc. These filters reduce the number of galaxies to 8896, with 4617 disc, 1592 \emph{G1}, and 1310 \emph{G2} galaxies. Here, it is important to note that the proportions of \emph{G1} and \emph{G2} galaxies are not significantly altered by the quality cuts performed above. Before the filters, the ratio of the number of these galaxies ($N_{\rm G1} / N_{\rm G2}$) was 1.24, and after removing 202 \emph{G1} galaxies and 141 \emph{G2} galaxies, the ratio $N_{\rm G1} / N_{\rm G2}$ equals 1.22. Thus, since there is no substantial change in the proportion, no bias related to imbalance of the two classes is introduced by cutting the sample in the way we describe here.

To compute $\langle \mu \rangle_e$ for the F814W filter, we retrieve the observed half-light radii ($R_{e \rm , F814W}$) and magnitudes ($m_{\rm F814W}$). On the other hand, to compute $\langle \mu \rangle_e$ for the rest-frame J-band, half-light radii ($R_{e, J}$) are obtained by evaluating the second-order Chebyshev polynomials using the coefficients provided by \cite{Nedkova2024}; however, we follow a different approach to obtain the rest-frame magnitudes. The dependence of half-light radius on wavelength is reasonably well-behaved due to the low order of the Chebyshev polynomials for this parameter. However, magnitudes are modelled without constraining the polynomial order in \cite{Nedkova2024}, making interpolation or extrapolation to redder wavelengths much less reliable. For this reason, we choose to retrieve rest-frame magnitudes ($m_{J}$) from a crossmatch with \cite{Santini2015}, more specifically, the J-band magnitudes from their \texttt{6a\_tau} model\footnote{This model corresponds to one of the various methods they use to estimate stellar masses. The "6a" prefix indicates the team and stellar template they used, while "tau" indicates that exponentially decreasing $\tau$ models have been assumed to parameterise the star formation history.}. The crossmatch is also performed using a 0.5 arcsec sky separation, resulting in a subsample of 2430 galaxies. Furthermore, this subsample is similarly cleaned by removing galaxies with large J-band radii ($> 30$ kpc), which yields 2024 objects: 400 \emph{G1} and 317 \emph{G2} galaxies. Because \cite{Santini2015} is restricted to the GOODS-S and UDS fields, this explains why only a fraction of galaxies in our sample can be used to fit the KR with rest-frame magnitudes. With all the necessary data, we perform orthogonal distance regression (ODR) to fit individual KR in redshift bins with a width of 0.2 and going from $z=0.2$ to $z=2.4$. We require at least 25 data points per fit in each redshift bin, and we estimate errors for the KR parameters using a bootstrap method with 1000 rounds. When referring to the relation being fitted in this work, we mean the following equation: $\langle \mu \rangle_{e} = a\,\log R_{e} + b$, where $a$ is the slope of the relation, $R_e$ is the half-light radius in units of kpc and $b$ is the intercept in units of $\rm{mag / arcsec^2}$. To illustrate, in Appendix \ref{app:KR} we show two examples of the KR fitting.

\subsection{Cluster Sample Crossmatch}
\label{sec:cluster_crossmatch}

We investigate the spatial distribution of bulge-dominated galaxies in clusters by combining our sample with a large cluster catalogue released by \cite{Wen2024}. Their catalogue is based on the DESI Legacy Imaging Surveys data \citep{Dey2019}, combined with spectroscopic information from other surveys when available. They identify over 1.58 million galaxy clusters by searching for stellar-mass overdensities in redshift slices around preselected massive galaxies. The galaxy cluster redshift distribution goes up to $z \sim 1.5$, and they provide physical properties such as the cluster radius ($r_{500}$) and mass ($M_{500}$). The galaxies in our sample are assigned to the closest cluster (in projected sky separation, $R_{\rm proj}$) within a redshift slice of $z_c \pm \Delta z$, with $\Delta z = 0.05(1+z)$, where $z_c$ and $z$ are the cluster and galaxy redshifts, respectively. Additionally, to avoid contamination of objects which are too far from the cluster centres, we only consider galaxies with $R_{\rm proj} \leq 3 r_{500}$. With these simple assignment criteria, our sample of sources assigned to clusters comprises 202 \emph{G1}, 431 \emph{G2} and 996 disc galaxies.

\subsection{Spin Parameter}
\label{sec:crossmatch_lambdaR}

In Section \ref{sec:rot_support}, we explore the rotational support in the galaxies of our sample through the use of the spin parameter ($\lambda_R$). The values of $\lambda_R$ reported in our results are derived from the cross-match between our sample and the catalogue released by \cite{MunozLopez2024}, which contains 106 galaxies located in the GOODS-S and COSMOS fields observed by various MUSE surveys. The sources are within the $0.1 \leq z \leq 0.8$ range and are selected to have a global signal-to-noise ratio $\geq 10$. We crossmatch the samples only in RA and DEC, using a maximum sky separation of 1 arcsec, yielding 17 disc, 11 \emph{G1}, and 18 \emph{G2} galaxies. We note that the values $\lambda_R$ and ellipticity ($\epsilon$) reported here are extracted from \cite{MunozLopez2026}, which are the corrected values from the original paper. In their original paper, when calculating the spin parameter, the systemic velocities of galaxies were not subtracted from the mean stellar velocity, resulting in an inaccurate estimation of the spin parameter. After the corrections, most galaxies have lower $\lambda_R$ values than previously reported. 

\section{Results}
\label{sec:results}

\subsection{Morphological Metrics}
\label{sec:morph_metrics}
\begin{figure}
     \centering
         \includegraphics[width=0.5\textwidth]{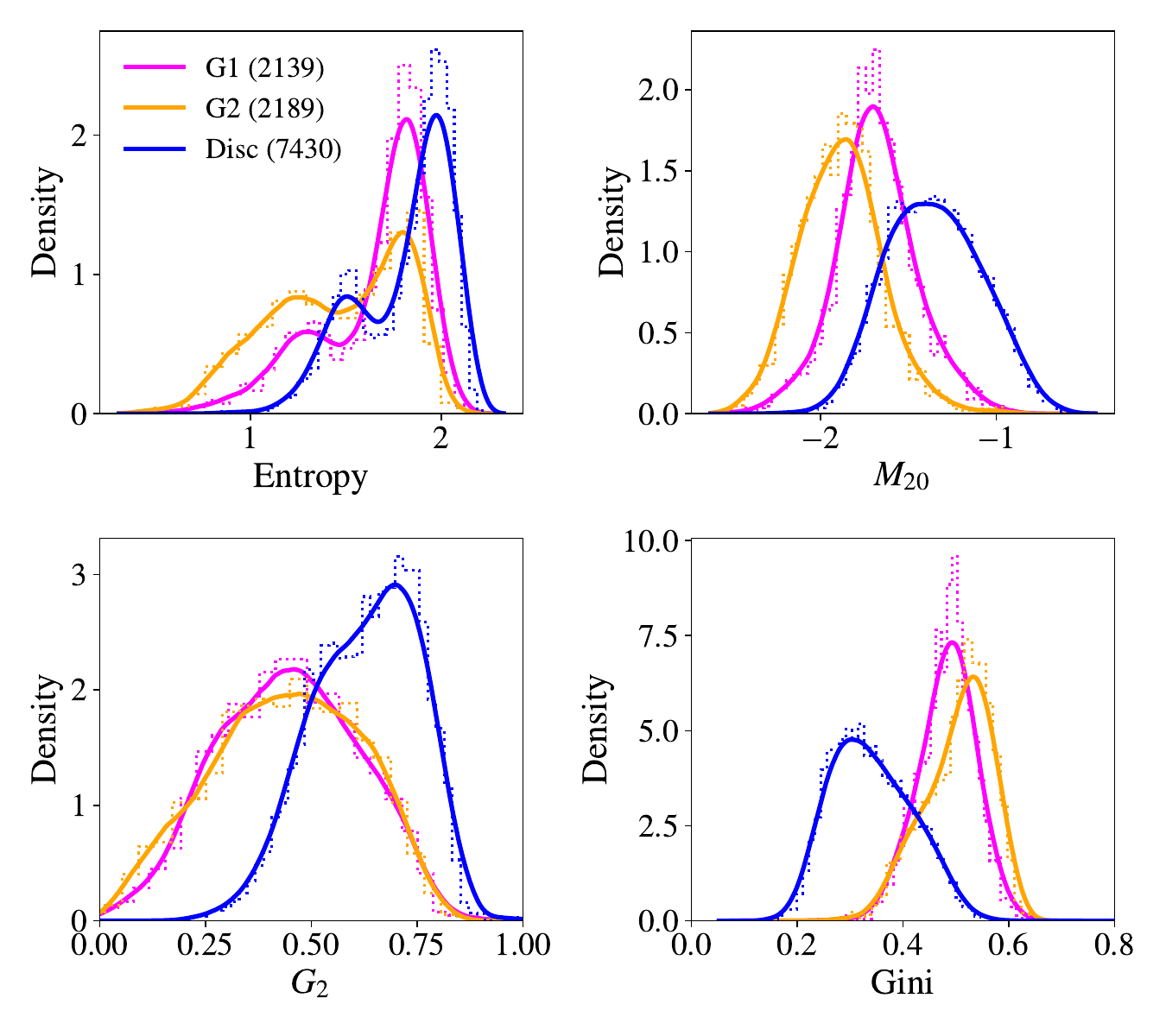}
     \caption{Distribution of metrics for galaxies classified as \emph{G1} (magenta), \emph{G2} (orange) and disc (blue). Solid lines indicate Epanechnikov kernel density estimations (KDEs) and dotted lines indicate histograms with bins defined by the Freedman-Diaconis rule. The vertical axes show probability density, so the area under the curves or histograms sums to 1. The bandwidth ($h_{\rm KDE}$) of the KDEs shown in all figures is defined according to an adaptation of Silverman's rule of thumb, where $h_{\rm KDE} = 2.34 \sigma_{r} N^{-1/5}$, where $N$ is the number of data points, and $\sigma_r = \min (\sigma, {\rm IQR}/1.349)$, with $\sigma$ being the standard deviation and IQR the interquartile range.}
     \label{fig:morph_metrics}
 \end{figure}

As presented in \cite{Sampaio2025}, the \emph{G1} and \emph{G2} families described here are defined as subpopulations of the spheroid class, using only bimodality in sSFR as the criterion. Since these two families of bulge-dominated galaxies are not \textit{directly} separated by morphological properties, it is important to understand the differences between them in this respect. In Figure \ref{fig:morph_metrics} we present the distributions of the MEGG metrics for discs, \emph{G1} and \emph{G2} galaxies. Except for the $G_2$ metric (the gradient pattern asymmetry), it is evident that \emph{G1} occupies an intermediate position between discs and \emph{G2} spheroids in all morphological metrics. The separation is more pronounced in $M_{20}$, indicating that \emph{G1} galaxies have a light distribution less centrally concentrated than \emph{G2} galaxies, but still more concentrated than disc galaxies. Although both families of bulge-dominated galaxies have similar distributions of the Gini coefficient, the \emph{G1} distribution peaks at a slightly lower value than \emph{G2}, again indicating that light is less concentrated in \emph{G1} galaxies. 
 
In Figure \ref{fig:morph_metrics}, we show all galaxies in our sample, and in an expanded figure in Appendix \ref{app:morph}, we show that the trend described here persists even in individual redshift bins (see Figure \ref{fig:morph_across_z}). This intermediate morphological status of \emph{G1} galaxies, between \emph{G2} and discs, may indicate that although their morphology is dominated by a spheroidal component, they retain structural characteristics of galaxies classified as discs.

\subsection{Structural Scaling Relations}
\label{sec:struct_relations}
\begin{figure}
	\centering
	\includegraphics[width=0.45\textwidth]{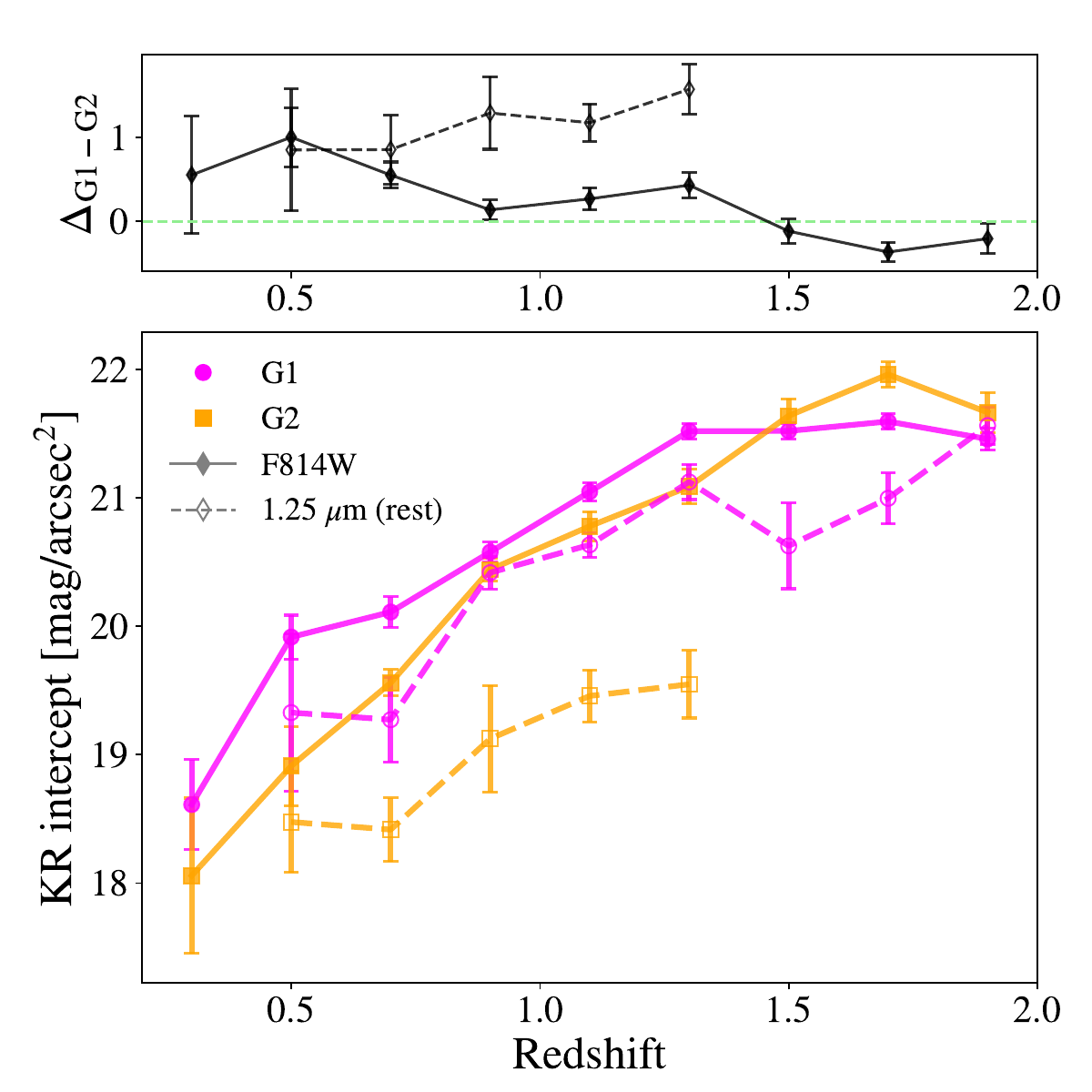}
	\caption{Intercept ($b$) of the Kormendy relations fitted for \emph{G1} (magenta) and \emph{G2} (orange) galaxies as a function of redshift. The lower panel shows the evolution of $b$ calculated from observed F814W (solid) and rest-frame J-band (dashed) magnitudes, with uncertainties estimated using 1000 bootstrap rounds of fitting, adopting the formula $Q_\sigma = 0.74(Q75 - Q25)$, where Q25 and Q25 are the 25th and 75th percentiles of $b$. The top panel shows the difference between \emph{G1} and \emph{G2} (black lines) in each redshift bin. In the top panel, the line styles of the black lines follow the line styles in the bottom panel, and the green dashed horizontal line indicates no difference between the intercepts. Error bars are estimated by propagation of uncertainty, addition in quadrature in this case.}
	\label{fig:kormendy}
\end{figure}

The difference between the two populations in the space of morphological metrics motivates further investigation into where each family lies on well-known scaling relations. The Kormendy relation \citep{Kormendy1977}, a projection of the fundamental plane \citep{DjorgovskiDavis1987, Dressler1987}, provides important physical information by linking the structural properties of early-type galaxies to their formation histories. Thus, we divide \emph{G1} and \emph{G2} galaxies into redshift bins and fit a KR for each family in each bin, using both observed magnitudes (from the F814W filter) and rest-frame magnitudes (J-band centred at 1.25 microns). The evolution of the KR intercept for each case is presented in Figure \ref{fig:kormendy}. The important result here is not the evolution of the intercept values of each family of bulge-dominated galaxies with redshift, but the difference between them across cosmic time. The comparison of the intercepts of the \emph{G1} and \emph{G2} families across redshift shows a clear trend: \emph{G1} galaxies have intercepts at lower surface brightness than \emph{G2} galaxies. This difference persists with redshift, regardless of the magnitudes used for the computation of $\langle \mu \rangle_e$. In fact, the fainter KR intercepts for \emph{G1} are even clearer if we consider the rest-frame 1.24 $\mu$m magnitude, instead of the observed F814W magnitude, with a difference larger than one standard deviation. At first glance, this distinction in mean surface brightness also implies a difference in stellar mass density between the two families, which could be related to different formation mechanisms or star-formation histories.

A potential concern when interpreting the evolution and offset of the Kormendy zero-point is the role of bandpass and stellar-population effects, since the intercept $b$ in $\langle \mu \rangle_{e} = a\,\log R_{e} + b$ is sensitive to the mapping between observed flux and rest-frame surface brightness. In our case, however, the comparison is explicitly differential: \emph{G1} and \emph{G2} are measured in the same imaging data and within the same redshift bins, so the dominant redshift-dependent terms (e.g. cosmological dimming and global photometric calibration) affect both families in a largely common way. 
Moreover, the separation between \emph{G1} and \emph{G2} is not defined by gross morphology but primarily by star-formation activity (via sSFR). A visual inspection of their images indicates that they are difficult to distinguish by eye (see Figure \ref{fig:examples}), consistent with the notion that the main body of the light distribution in both populations is governed by a spheroid component. 
This motivates the interpretation that the persistent offset in $b$ reflects, to first order, a genuine structural difference, i.e., at fixed $R_{e}$, \emph{G2} systems occupy brighter $\langle \mu \rangle_e$, plausibly tracing higher central stellar densities and more centrally concentrated profiles. At the same time, we do not require the stellar populations of \emph{G1} and \emph{G2} to be identical: even modest residual or recently quenched star formation in \emph{G1} can alter mass-to-light ratios and colour gradients, and thus can contribute non-negligibly to the observed $\langle \mu \rangle_{e}$ in a redshift-dependent manner. The key point is therefore not that population-driven corrections vanish, but that they are expected to be sub-dominant compared to the structural signal implied by a stable, family-dependent offset observed among otherwise similar bulge-dominated galaxies. Accordingly, we treat stellar-population differences as a secondary contribution that may modulate, but is unlikely to fully drive, the observed \emph{G1}-\emph{G2} separation in the zero-point of the Kormendy relation. 

\subsection{Rotational Support}
\label{sec:rot_support}

\begin{figure}
    \centering
\includegraphics[width=0.49\textwidth]{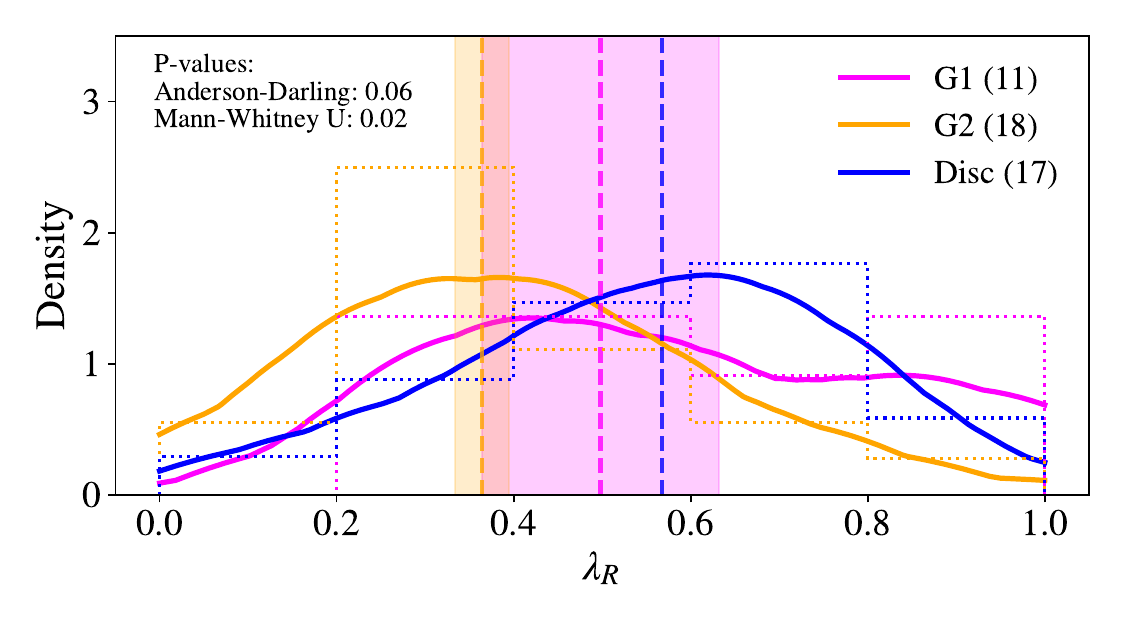}
    \caption[caption]{Distribution of the spin parameter ($\lambda_R$) for galaxies in crossmatch with sample from {\cite{MunozLopez2024}}. Solid lines indicate Epanechnikov KDEs and dotted lines indicate histograms with bins defined by the Freedman-Diaconis rule. The vertical axes show probability density, so the area under the curves or histograms sums to 1. Vertical dashed lines indicate median values of $\lambda_R$ and shaded areas show the uncertainties of the medians estimated using 1000 bootstrap rounds, adopting the formula $Q_\sigma = 0.74(Q75 - Q25)$, where Q25 and Q25 are the 25th and 75th percentiles of $\lambda_R$. In the top-left corner, we present the p-values of two-sample statistical tests performed using \emph{G1} and \emph{G2} samples.}
    \label{fig:lambdaR}
\end{figure}

The analysis of morphology and KR presented so far supports a scenario in which \emph{G1} and \emph{G2} differ in structure. We next examine the kinematics of both bulge-dominated families, as these properties hold key insights into the different formation pathways of early-type galaxies. The spin parameter $\lambda_R$ can be used to quantify rotational support in early-type galaxies, and it has already been measured for a few sources up to intermediate redshifts \citep{MunozLopez2024, Mozumdar2025b}. Thus, we cross-match our sample with the catalogue from \cite{MunozLopez2024}, which has MUSE observations in the redshift range $0.1 < z < 0.8$. 
In Figure \ref{fig:lambdaR} we show the distribution of the spin parameter for \emph{G1}, \emph{G2} and disc\footnote{Classified as discs according to our pipeline, and not using any classification scheme from \cite{MunozLopez2024}.} galaxies, alongside the median values of $\lambda_R$ for each sample. We found that the \emph{G1} sample tends to have higher values of spin than \emph{G2}, with the offset between the $\lambda_R$ medians being 0.14, corresponding to a difference of 0.15 in $V/\sigma$ (assuming the relation from \cite{Emsellem2007}). Here, we emphasise that the median error of the $\lambda_R$ values shown is 0.014, and that the low-number statistics dominate the uncertainty in our results. We quantify the uncertainties in the $\lambda_R$ medians for the \emph{G1} and \emph{G2} samples using the $Q_\sigma$ estimator across 1000 bootstrap runs, which only provides a conservative estimate of their variation. Although the error bars for the medians are not negligible, the trend of \emph{G1} having higher rotational support than \emph{G2} is already evident even with a small number of points.

In order to quantify the significance of the difference between the median $\lambda_R$ of both spheroid families, we performed two statistical tests, the Anderson-Darling 2-sample test and the Mann-Whitney U rank test, with the p-values shown in Figure \ref{fig:lambdaR}. The Mann–Whitney test indicates a statistically significant shift between the two populations (p=0.02), whereas the Anderson–Darling test does not reject the null hypothesis of identical overall distributions at the 5\% level (p=0.06). This suggests that the main difference may be due to the location of the central tendency rather than a major change in the overall distribution shape. Additionally, a second cross-match with the MAGNUS sample \citep{Mozumdar2025b} yields 6 galaxies, all of which are of the \emph{G2} population. The median value of the intrinsic $\lambda_R$ for these 6 objects is $\sim 0.23$, which is closer to the median of the 15 \emph{G2} presented in Figure \ref{fig:lambdaR}.
Unfortunately, the cross-match with \cite{MunozLopez2024} data yields only a few galaxies for comparison, and the results for $\lambda_R$ should be interpreted with caution. Nonetheless, the distribution of this parameter in Figure \ref{fig:lambdaR} suggests that \emph{G1} galaxies have higher rotational support.

One could argue that, given the lower stellar mass and fainter surface brightness of \emph{G1} galaxies, kinematic measurements are likely more challenging for \emph{G1} than for \emph{G2}, so the \emph{G1} sample with $\lambda_R$ measurements may be biased toward higher-spin galaxies. However, we performed a linear correlation analysis of the 106 galaxies from the \cite{MunozLopez2024} catalogue, looking for dependences between $\lambda_R$ and other physical and observational properties (taken from their Table B.1), such as stellar mass, F160W band magnitude, signal-to-noise ratio (SNR) and effective radius. We found that none of these quantities shows a significant correlation (coefficient of determination $<$ 0.1) with the spin parameter. Thus, in the context of the sample considered here, we do not expect differences in stellar mass, size, magnitude, or SNR between \emph{G1} and \emph{G2} to affect their measures of $\lambda_R$. Furthermore, a potential concern in interpreting our results for $\lambda_R$ is its variation with radius. As we are using an averaged global estimate of the projected spin, differences in the brightness profiles of \emph{G1} and \emph{G2} could affect how representative the $\lambda_R$ values are for the galaxies' central regions. However, as reported by \cite{MunozLopez2024} (see their section 4.3 and figure 4), the variation of $\lambda_R$ with the aperture in which it is measured is not large for most galaxies in the sample. Thus, we do not expect that the kinematics of the outskirts are driving higher values of $\lambda_R$ in the \emph{G1}sample, for example. Nonetheless, a more detailed analysis of a larger sample of \emph{G1} and \emph{G2} galaxies and their $\lambda_R$ is required to completely remove any ambiguity from using global values of the spin parameter.

What makes this result particularly significant is that the separation in $\lambda_R$ does not simply reproduce the morphological classification in another form. Although \emph{G1} and \emph{G2} are both classified as bulge-dominated galaxies, they do not occupy the same kinematic domain. \emph{G1} overlaps almost entirely with the discs (classified here according to our pipeline\footnote{It is important to note that \emph{G1} galaxies have different morphologies than disc galaxies indeed, as we show in Appendix \ref{app:purity_contamination} that on average only 8\% of the spheroid galaxy sample may be contaminated with discs.}, independently of any morphological assignment by \citealt{MunozLopez2024}), whereas \emph{G2} is displaced toward lower $\lambda_R$, in the direction expected for a dynamically hotter population with a smaller contribution from ordered stellar motions. The most immediate interpretation of this result is within the now-standard fast-/slow-rotator framework for early-type galaxies. Since $\lambda_R$ was introduced precisely to quantify projected specific angular momentum in early-type galaxies, the fact that the two bulge-dominated families separate in this parameter indicates that the \emph{G1}/\emph{G2} bifurcation has a genuine dynamical meaning. In this sense, \emph{G1} appears as the branch more naturally associated with the fast-rotator side of the spheroid population, while \emph{G2} is shifted toward the lower-$\lambda_R$, more dispersion-dominated regime conventionally associated with dynamically hotter and quenched systems. Despite the difference in $\lambda_R$, almost all \emph{G1} and \emph{G2} galaxies present in our crossmatch are classified as fast-rotators. Thus, this does not mean that \emph{G1} should be reclassified as a disc population, nor that \emph{G2} must be identified one-to-one with the local slow-rotator class in the context of ATLAS$^{\rm 3D}$ sample \citep{Cappellari2011}. The point is more fundamental and more important: spheroidal morphology does not uniquely define an internal dynamical state. Galaxies that look similar at the level of global structure in photometry may nevertheless retain substantially different fractions of ordered stellar motion, and it is precisely this mismatch between morphology and kinematics that gives physical meaning to the distinction between \emph{G1} and \emph{G2}. Under this interpretation, the bifurcation is not merely a secondary consequence of differences in stellar mass or star-formation activity, but evidence that the spheroid population itself is split between at least two assembly channels, one in which angular momentum is more efficiently preserved and another in which it is more effectively reduced by heating, mergers, or repeated perturbative evolution.

Once this interpretation is established, a tentative indirect connection between \emph{G1} galaxies and pseudo-bulges becomes possible, but only at a secondary evolutionary level. Pseudo-bulges are usually defined as central structures embedded in disc galaxies and recognised through diagnostics tied specifically to the inner component, including rotational support, flattening, nuclear structure, and central star formation. The \emph{G1} family presented here emerges from the bimodality in sSFR within a sample already classified as bulge-dominated at the global level, that is, from a distinction in star formation activity rather than from a decomposition of the central light distribution. For this reason, \emph{G1} should not be directly identified as equivalent to the pseudo-bulge population itself, either conceptually or observationally. Even so, the kinematic behaviour of \emph{G1} enables a novel interpretation. If a bulge-dominated galaxy population can retain $\lambda_R$ values comparable to those of discs, then at least one route to bulge growth must allow a substantial fraction of the angular momentum of the progenitor system to survive morphological transformation. That is exactly the kind of physical continuity one would expect if some present-day pseudo-bulges descend from evolutionary channels in which disc kinematics is not erased but only reworked. Therefore, the \emph{G1} population may trace a mode of spheroid evolution that remains more strongly coupled to the structural imprint of discs, thereby offering a plausible antecedent for at least part of the pseudo-bulge phenomenon at low redshift.

\subsection{Environmental Dependence}

\begin{figure}
	\centering
	\includegraphics[width=0.5\textwidth]{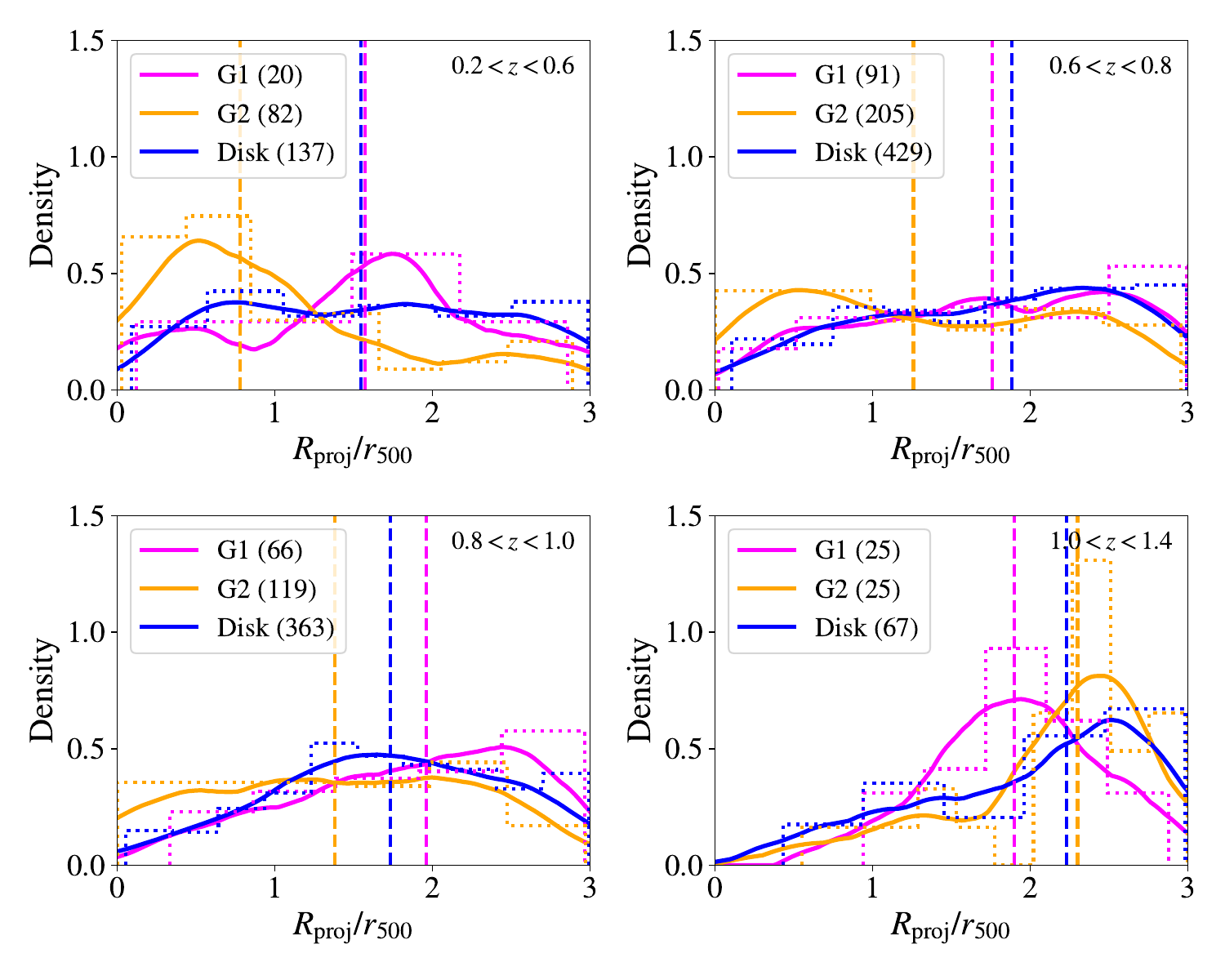}
	\caption{Distribution of projected clustercentric distances for \emph{G1} (magenta), \emph{G2} (orange) and disc (blue) galaxies. Solid lines indicate Epanechnikov KDEs and dotted lines indicate histograms with bins defined by the Freedman-Diaconis rule. The vertical axes show probability density, so the area under the curves or histograms sums to 1. Vertical dashed lines indicate median values for each class. The redshift intervals on the panels are chosen to optimise the number of galaxies in each bin.}
	\label{fig:clusters}
\end{figure}

It has long been known that the morphology of galaxies is related to their environmental density \citep{Dressler1980}, with early-type galaxies being more abundant in denser environments. Specifically within clusters, it is observed that the fraction of late-type galaxies decreases toward the cluster centres \citep{Whitmore1993,Fasano2015}. To explore whether the populations of bulge-dominated galaxies analysed here occupy different loci within the cluster environment, we combine our sample with the cluster catalogue (see Section \ref{sec:cluster_crossmatch}) from \cite{Wen2024}. In Figure \ref{fig:clusters} we present the projected cluster-centric distances ($R_{\rm proj}$) scaled by the $r_{500}$ of their respective host clusters. The comparison between the $R_{\rm proj}/r_{500}$ distributions of the bulge-dominated galaxy families reveals a tendency of \emph{G1} galaxies to be farther from the cluster centre than \emph{G2} galaxies. As we move to higher redshifts, the distributions of \emph{G1} and \emph{G2} become more similar, especially at $1 < z < 1.4$, where both families are mostly located at $R_{\rm proj} > r_{500}$.

 Although the number of galaxies presented in Figure \ref{fig:clusters} is only a fraction of our sample of bulge-dominated galaxies, we can observe a clear trend in the relative spatial distribution of \emph{G1} and \emph{G2} galaxies within clusters at the redshift interval spanning $0.2 < z < 1$. The fact that \emph{G1} galaxies are more prevalent in the outer parts of clusters suggests that these galaxies can maintain star formation precisely because they have not yet been quenched by the cluster environment. Because they lie at the edges of clusters, their cold gas reservoirs were not affected by interactions with the intracluster medium (e.g., ram-pressure stripping), nor were they harassed or cannibalised by other galaxies. At the same time, in this scenario, \emph{G2} galaxies represent a channel for the evolution of more massive bulge-dominated galaxies, which cease star formation first and contribute to the buildup of higher galaxy number densities in the inner parts of the clusters observed today. In parallel, it is worth noting that \emph{G1} galaxies follow a similar trend to discs, which are also mainly outside 1 $r_{500}$, reinforcing the interpretation given above. It is important to emphasise here that \emph{G1} galaxies are not misclassified disc galaxies, but bulge-dominated galaxies with star-formation activity similar to discs. As the tests in \cite{Kolesnikov2025} show, the contamination of discs in the spheroid class is generally below $\sim 15$\% at $z\leq1.4$.
 
Although the galaxy number count is not large in each redshift bin, given the limited data available, we can already detect an important trend in the environment of the bulge-dominated galaxy families. The fact that the more star-forming bulge-dominated galaxies lie mostly on the outskirts of clusters suggests that these objects may have undergone a quieter evolution than the \emph{G2} galaxies at the same redshift. 

\section{Discussion}
\label{sec:discussion}
\subsection{Evolution channels of bulge-dominated galaxies}

\begin{figure*}
	\centering
	\includegraphics[width=0.99\textwidth]{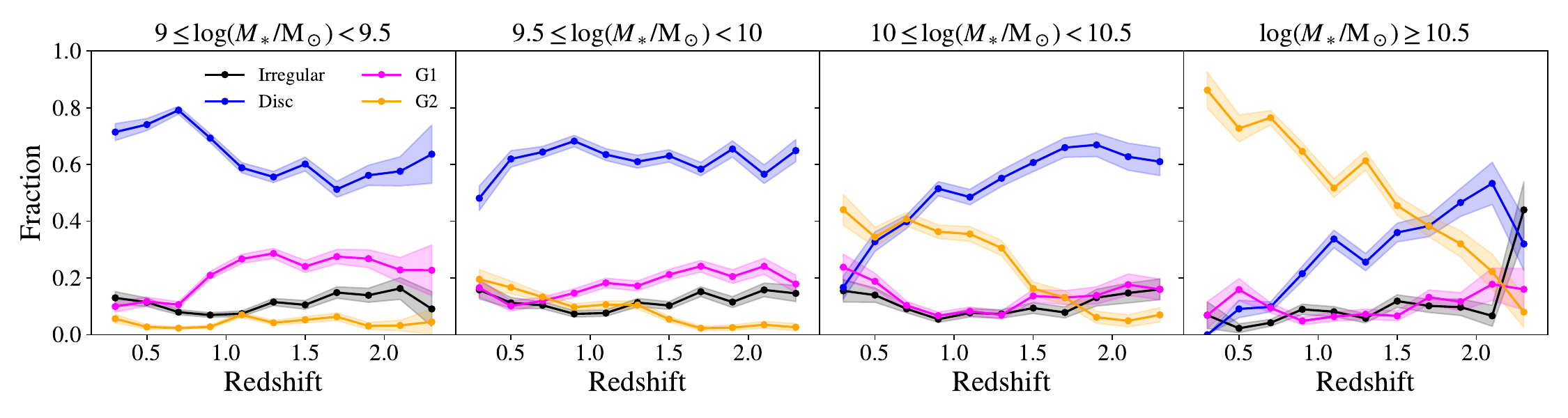}
	\caption{Fraction of discs (blue), \emph{G1} (magenta), \emph{G2} (orange) and irregular (grey) galaxies as a function of redshift in different stellar mass bins. The uncertainties shown as shaded regions are calculated through a multinomial distribution.}
	\label{fig:fractions}
\end{figure*}

The distinction between \emph{G1} and \emph{G2} galaxies in morphology, mean surface brightness, rotational support, and environment paints a general picture in which these two populations of bulge-dominated galaxies follow different evolutionary paths. This motivates us to investigate the evolution of each morphological type in our sample, separating galaxies by stellar mass to identify trends across different mass regimes. In Figure \ref{fig:fractions}, we present the evolution of the fraction of \emph{G1}, \emph{G2}, discs and irregular galaxies, with the fractions being relative to the total number of galaxies in our CANDELS sample at a given redshift and stellar mass bin. If we focus on high-mass galaxies ($M_\ast > 10^{10} \ \rm M_\odot$), there is a clear anti-correlation between the fractions of discs and \emph{G2} galaxies, with the latter increasing towards lower redshifts, while the former decreases, apparently at a similar rate. Meanwhile, the \emph{G1} and irregular fractions remain mostly constant in the same redshift interval, being subdominant, with values around $\sim 15 \%$ each. Since these are fractions of different morphological types within fixed stellar-mass bins at different redshifts, rather than a measurement that tracks individual galaxies or their progenitors, this anti-correlation cannot by itself establish a direct evolutionary link between the two populations. It could also reflect, for example, progenitor bias or the entry of new galaxies into the mass bins through growth or accretion. Bearing this caveat in mind, the trend is consistent with a scenario in which part of the more massive disc galaxies are being destroyed to form \emph{G2} galaxies, whereas \emph{G1} galaxies remain mostly a roughly stable population of lower-mass, bulge-dominated galaxies. The picture changes when we analyse low-mass galaxies ($10^9 \ {\rm M_\odot} < M_\ast < 10^{10} \ \rm M_\odot$), where the disc morphology remains dominant across the entire redshift interval shown, and \emph{G1} is the most common type of bulge-dominated galaxy, although with fractions remaining below 30\%.
The higher \emph{G1} fractions relative to \emph{G2} at lower masses are consistent with \emph{G1} tracing a population of lower-mass disc galaxies that are already building up their central spheroidal components, though confirming this connection would require complementary evidence, such as number densities, rather than fractions alone.

As shown in \cite{Sampaio2025}, \emph{G1} galaxies are, on average, always less massive than \emph{G2} (see their figure 6). Given the evolution of these two families in Figure \ref{fig:fractions}, one possible interpretation is that the emergence of the \emph{G2} population is associated with the quenching of more massive star-forming disc galaxies, potentially driven by mergers and environmental influence, while \emph{G1} galaxies follow a distinct evolutionary pathway more closely related to bulge growth in disc galaxies. This interpretation is further supported by the fact that \emph{G1} galaxies are closer to the disc population in morphological and kinematic terms, as explored in Section \ref{sec:results}. Additionally, the difference in KR offsets indicates that the \emph{G1} population has lower surface brightness than \emph{G2}, implying a lower stellar mass concentration at the centres of \emph{G1} galaxies. Previous works have already suggested that the growth of a dense stellar core can precede star-formation quenching \citep{Cheung2012,Fang2013,vanDokkum2014}, with the stellar surface density at the inner kpc ($\Sigma_1$) serving as a cosmic clock to trace the evolution of galaxies \citep{Barro2017,Chen2020,EstradaCarpenter2020}. The fact that the \emph{G2} branch has a brighter $\langle \mu \rangle_e$ at fixed size and is the bulge-dominated family with lower sSFR is consistent with this population having undergone a more rapid transition towards quiescence than \emph{G1}, which remains predominantly star-forming. Taken together, while Figure 6 alone cannot establish a direct evolutionary link, the convergence of these independent lines of evidence (structural, kinematic, and morphological) supports this scenario.

\subsection{\emph{G1}-\emph{G2} bifurcation and the emergence of pseudo-bulges}
A possible interpretation of the \emph{G1}-\emph{G2} bifurcation is that it is the higher-redshift analogue of the long-recognised separation between pseudo-bulges and classical bulges in the nearby Universe, but the mapping should be stated as a hypothesis rather than a complete identification for all galaxies in both families. In the low-$z$ literature, pseudo-bulges are usually associated with disc-linked growth: they tend to remain connected to ongoing star formation, exhibit lower central densities at fixed size, and deviate from the scaling relations followed by spheroids assembled through more dissipative or violent channels \citep[e.g.][]{KormendyKennicutt2004,FisherDrory2016}. Our classification is consistent with this logic in a controlled way: \emph{G1} and \emph{G2} are both bulge-dominated in morphology and difficult to separate by eye, yet they occupy distinct loci in non-parametric metric space and in the KR plane, with \emph{G2} systematically brighter in $\langle\mu\rangle_e$ at fixed $R_e$, suggesting higher stellar mass surface density, while \emph{G1} retains higher sSFR and fainter surface brightness. If the dominant stellar body in both families is dispersion-dominated, then the persistence of the Kormendy zero-point offset shown in Figure \ref{fig:kormendy} can be read primarily as a structural statement. In this sense, \emph{G1} becomes a plausible pseudo-bulge candidate population at intermediate redshift: not because it looks disky visually, but because it appears to be a bulge-dominated system whose structure and star-formation state remain closer to the star-forming disc galaxy population. The idea that spheroid scaling relations can already be established early, and can be measured and compared across epochs, has recently been emphasised in the JWST context by \citet{BorgohainSaha2026}, who discuss the emergence of a Kormendy-like relation at very high redshift; our result is complementary in that it points to a bifurcation within the bulge-dominated population once the relation is studied with sufficient dynamic range. The immediate observational consequence is testable: if \emph{G1} truly connects to the pseudo-bulges at low $z$, it should show, at fixed stellar mass, systematically lower central stellar-mass surface density (or $\Sigma_1$), stronger rotational support, and a tighter coupling to disc indicators than \emph{G2}, whereas \emph{G2} should track the classical-bulge direction in density and quenching diagnostics \citep[e.g.][]{KormendyKennicutt2004,FisherDrory2016}. We therefore treat the \emph{G1}-\emph{G2} split as a candidate early manifestation of the pseudo/classical bulge divergence. 

Having established the observational case for the connection of \emph{G1} galaxies with pseudo-bulges, it is equally important to define the limits of that association. We emphasise here that \emph{G1} galaxies should not be identified with the pseudo-bulge population itself, and the distinction is important both conceptually and observationally. An important caveat is that pseudo-bulges are defined in the local Universe as central structures embedded in disc galaxies, usually recognised by a combination of properties indicating a strong connection to disc-driven evolution, such as rotational support, flattened morphology, ongoing or recent central star formation, nuclear bars, rings, or other signatures of secular inflow. By contrast, the \emph{G1} family in our analysis is not isolated through any direct bulge diagnostic of that kind. It emerged from the distribution of sSFR within a sample already classified globally as bulge-dominated, i.e., from a separation in stellar population activity rather than from a decomposition of the central component. This difference in construction matters. A pseudo-bulge is a statement about the physical nature of the inner structure of a disc galaxy; \emph{G1} is, at this stage, a statement about a family of bulge-dominated systems whose star formation history differs from that of the more quiescent \emph{G2} branch. The two classes may overlap in physical processes behind their formation, but they are not equivalent populations of galaxies. In particular, nothing in the present definition of \emph{G1} guarantees that these galaxies host a discy central component in the strict sense required by the pseudo-bulge framework, nor that their inner light profile, flattening, or kinematics would meet the usual criteria used for nearby galaxies. The safer interpretation is therefore evolutionary rather than taxonomic: \emph{G1} may trace a channel of bulge growth that remained more strongly coupled to the evolution of a stellar disc, and therefore it may represent a plausible antecedent of at least part of the pseudo-bulge population seen at low redshift.
Nonetheless, we stress the notion that calling \emph{G1} itself a pseudo-bulge population would collapse two distinct levels of description into one. One concerns the present-day structural identity of a central component, the other concerns the broader evolutionary path of galaxies already globally classified as bulge-dominated. Thus, the nature of the connection between \emph{G1} and pseudo-bulges made here is treated more as a hypothesis to be explored in the future.

\section{Conclusions}
\label{sec:conclusions}

In this work, we analyse the morphological, structural, kinematic, and environmental properties of bulge-dominated galaxies selected from CANDELS, building upon the \emph{G1}/\emph{G2} bimodality in specific star formation rate originally identified by \cite{Sampaio2025}. Our main findings can be summarised as follows:

\begin{itemize}
    \item The two families of bulge-dominated galaxies, \emph{G1} and \emph{G2}, occupy systematically different regions in non-parametric morphological metric space. In M20, Gini, Entropy, and \emph{G2}, \emph{G1} galaxies consistently occupy an intermediate position between disc galaxies and \emph{G2} spheroids. This intermediate structural character persists across $0.2 < z < 1.4$ for some metrics, indicating that \emph{G1} galaxies, despite being morphologically bulge-dominated, retain structural characteristics associated with the disc population.
    
    \item The analysis of the Kormendy relation reveals a persistent offset in the zero-point between the two families across the wide redshift range ($0.2 \leq z < 1.4$). \emph{G2} galaxies systematically exhibit brighter mean effective surface brightness at fixed effective radius, a result that holds whether observed F814W or rest-frame J-band magnitudes are used. We interpret this offset primarily as a structural signal: \emph{G2} galaxies harbour higher central stellar densities and more concentrated profiles than \emph{G1}, consistent with a relatively more rapid evolutionary pathway toward quiescence.
    
    \item A comparison between 29 galaxies with IFS observations shows that \emph{G1} galaxies have systematically higher spin parameters than \emph{G2}, with a difference of 0.14 in their medians. Although both families are predominantly fast rotators, this kinematic distinction may indicate that the \emph{G1}/\emph{G2} bifurcation corresponds to two distinct assembly channels: one in which angular momentum is more efficiently preserved, and one in which it is more effectively reduced by mergers, heating, or repeated perturbative evolution. Future analysis of larger samples of \emph{G1} and \emph{G2} galaxies with IFS observations will be crucial to further reinforce this proposed scenario.
    
    \item The analysis of the projected clustercentric distances reveals that \emph{G1} and \emph{G2} galaxies occupy different loci within the cluster environment. Across the redshift interval $0.2 < z < 1.0$, \emph{G1} galaxies are preferentially found in the outskirts of clusters (large $R_{\rm proj}/r_{500}$), while \emph{G2} galaxies are more concentrated closer to the cluster inner regions. This trend suggests that \emph{G1} galaxies have not yet experienced cluster-driven quenching mechanisms such as ram-pressure stripping and harassment, allowing them to sustain star formation. The convergence of the two distributions at $1.0 < z < 1.4$ is consistent with a picture in which the cluster environment progressively amplifies the bimodality toward lower redshifts. 
    
    \item The redshift evolution of morphological fractions, separated by stellar mass, shows that at high masses ($M_\ast > 10^{10} \rm M_\odot$) the \emph{G2} fraction increases as the disc fraction decreases toward lower redshift - a trend consistent with, though not direct evidence for, a transformation channel from discs to \emph{G2} galaxies - while \emph{G1} remains a subdominant but stable population. At lower masses ($10^9 \ {\rm M_\odot} < M_\ast < 10^{10} \ \rm M_\odot$), \emph{G1} is the dominant bulge-dominated type and discs remain prevalent, a pattern consistent with \emph{G1} tracing a channel of gradual bulge growth in lower-mass disc galaxies rather than a signature of dissipative quenching events. However, these trends reflect fractions within fixed stellar-mass bins, not individual galaxies tracked over time, and should be read as suggestive rather than conclusive evidence of an evolutionary connection.
    
    \item Finally, these results suggest that the \emph{G1}/\emph{G2} bifurcation may be a plausible early manifestation of the classical/pseudo-bulge dichotomy observed in the nearby Universe. \emph{G1} galaxies share with pseudo-bulges a closer structural and kinematic coupling to the disc population and fainter surface brightness at fixed size. However, \emph{G1} should not be directly identified as pseudo-bulges, as the two categories are defined through different observational diagnostics. The safer interpretation is that \emph{G1} galaxies may trace a mode of spheroid assembly in which the angular momentum and structural memory of a disc progenitor are partially preserved, providing a physical connection to at least part of the pseudo-bulge population seen at $z \sim 0$.
\end{itemize}

Future progress will require integral-field spectroscopy of larger, statistically complete samples spanning the full redshift range $0.2 < z < 2.4$, expanded environmental characterisation beyond the cluster regime explored here, and spatially resolved stellar population modelling capable of disentangling the star formation histories and internal structural properties of individual bulge-dominated galaxies. Facilities such as the JWST, the Euclid telescope, the Nancy Grace Roman Space Telescope, and the Extremely Large Telescope will be decisive in this effort, providing the combination of spatial resolution and sensitivity needed to extend kinematic and photometric diagnostics of the bulge-dominated bimodality to the epochs when the divergence between these two evolutionary channels is first established.

\section*{Acknowledgements}
The authors thank the referee for the comments that led to an improved version of the manuscript. RFF and RRdC acknowledge the support from FAPESP through the grant 2025/24207-0. VMS acknowledges support from ESO through grant ORP026/2021, and CLD from the ESO Comité Mixto through grant ORP037/2022. VMS also thanks the financial support from the Agencia Nacional de Investigación y Desarrollo (ANID) through the Millennium Science Initiative Program NCN2024\_112. RFF acknowledges the use of
AI language models (ChatGPT) and Grammarly for assistance in improving the manuscript's clarity and readability. All scientific analyses, interpretations, and
conclusions were developed and verified by the authors.

\section*{Data Availability}
The data underlying this paper originate from the CANDELS database and the works from \cite{Kolesnikov2025}, \cite{Nedkova2024}, \cite{Wen2024}, \cite{MunozLopez2026} and \cite{Mozumdar2025b}. The data underlying this article will be shared upon reasonable request to the corresponding author.

\bibliographystyle{mnras}
\bibliography{paper.bib}

\appendix

\section{Purity and contamination of morphological classes}
\label{app:purity_contamination}

\begin{figure}
    \centering
    \includegraphics[width=0.4\textwidth]{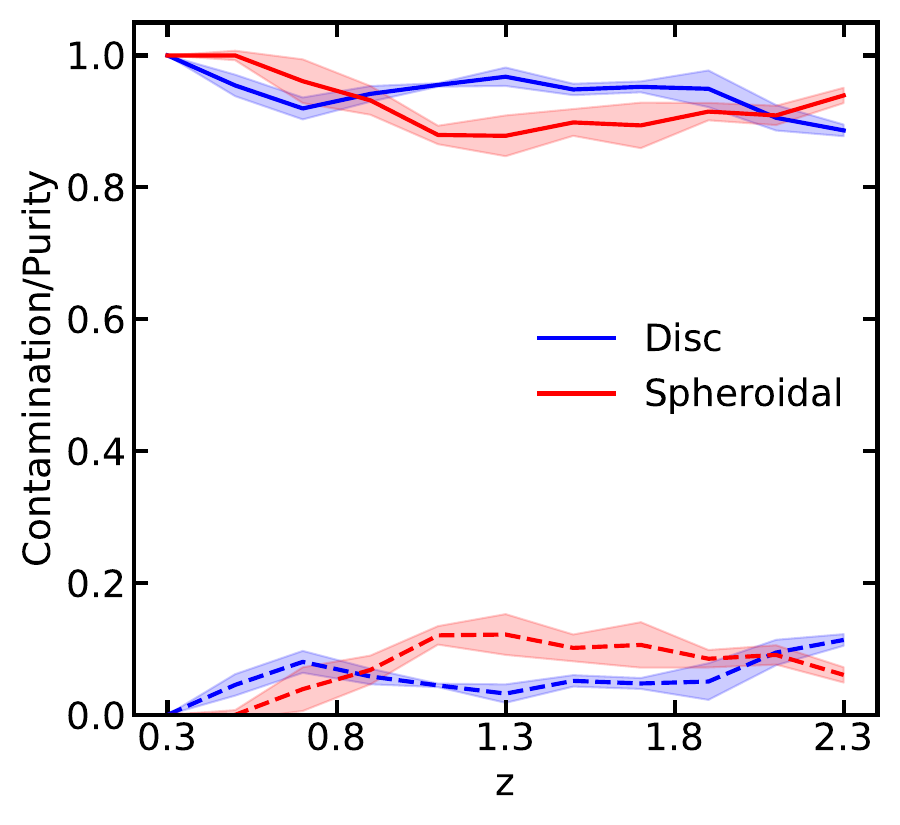}
    \caption[caption]{Contamination and purity of spheroid (red) and disc (blue) classification. This figure uses the same data as Figure 9 in \cite{Kolesnikov2025} but computes contamination (dashed lines) and purity (solid lines) rather than fractions.}
    \label{fig:purity}
\end{figure}

In this appendix, we present the expected contamination of discs into spheroids and vice versa according to results from \cite{Kolesnikov2025}. According to the results presented in Section \ref{sec:results}, \emph{G1} galaxies seem to occupy an intermediate position between \emph{G2} galaxies and disc galaxies. One could argue that the status of the \emph{G1} sample as an intermediate population of galaxies is driven by contamination from discs into this subsample of bulge-dominated galaxies. To address this potential issue, we present the purity\footnote{Also called precision in machine learning contexts.} metric ($P$), computed as the ratio $P = {\rm TP / (TP + FP)}$, where TP and FP stand for true positives and false positives, respectively. We also present the complementary quantity, the contamination, which is computed as 1 - $P$. Following \cite{Kolesnikov2025}, here each galaxy classification in the lowest redshift bin ($0.2 \leq z < 0.4$) is treated as the ground truth and compared with the classifications obtained from images of the same galaxy degraded with the \textsc{FERENGI} code. Our estimates of purity and contamination as a function of redshift are presented in Figure \ref{fig:purity}. The average contamination of discs into spheroids is 8\% and of spheroids into discs is 6\%. Furthermore, the contamination was examined across the entire redshift range, comprising 10 redshift bins (each with an ensemble of 100 models). In other words, for each of these 10 bins, \cite{Kolesnikov2025} obtained purity and contamination when comparing the assumed ground truth and the prediction provided by the ensemble of the tested bin. These low contamination percentages are expected, given the pipeline design. In the CNN step of the classification pipeline, we do not rely solely on a single CNN but on an ensemble of 100 CNNs. Each network is trained on identical data but with different random selections of the training and validation sets, introducing variation and making the predictions more robust. The predictions from the 100 CNNs yield a probability distribution with a peak near 0 and another near 1, as shown in Figure 5 of \cite{Kolesnikov2025}. Galaxies with probabilities below 0.1 are identified as discs, and the ones with probabilities above 0.9 are identified as spheroids. Thus, since our classification relies on the probability distribution generated by multiple CNNs, if the networks exhibit large disagreement in their predictions, the probability assigned to the classification will be closer to 0.5, and the galaxy will fall into the irregular class. This reduces the likelihood of significant contamination between classes (from one peak to the other).

As we mentioned in Section \ref{sec:galaxy_sample}, the separation of \emph{G1} and \emph{G2} spheroids is created based simply on the sSFR bimodality of galaxies classified as spheroids. Here, we emphasise that classification is performed first, and only afterwards are the \emph{G1} and \emph{G2} samples created based solely on star-formation activity. Thus, in our method, there is no specific selection or separation that directly divides bulge-dominated galaxies by morphology.

\section{Kormendy relation fitting}
\label{app:KR}

\begin{figure}
    \centering
    \includegraphics[width=0.45\textwidth]{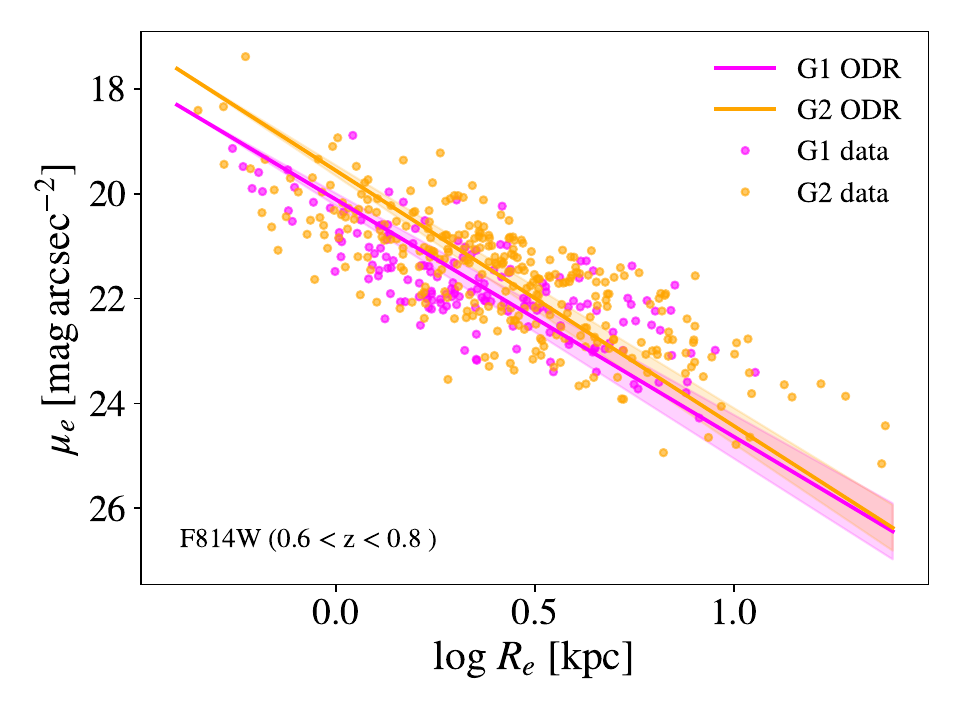}
    \includegraphics[width=0.45\textwidth]{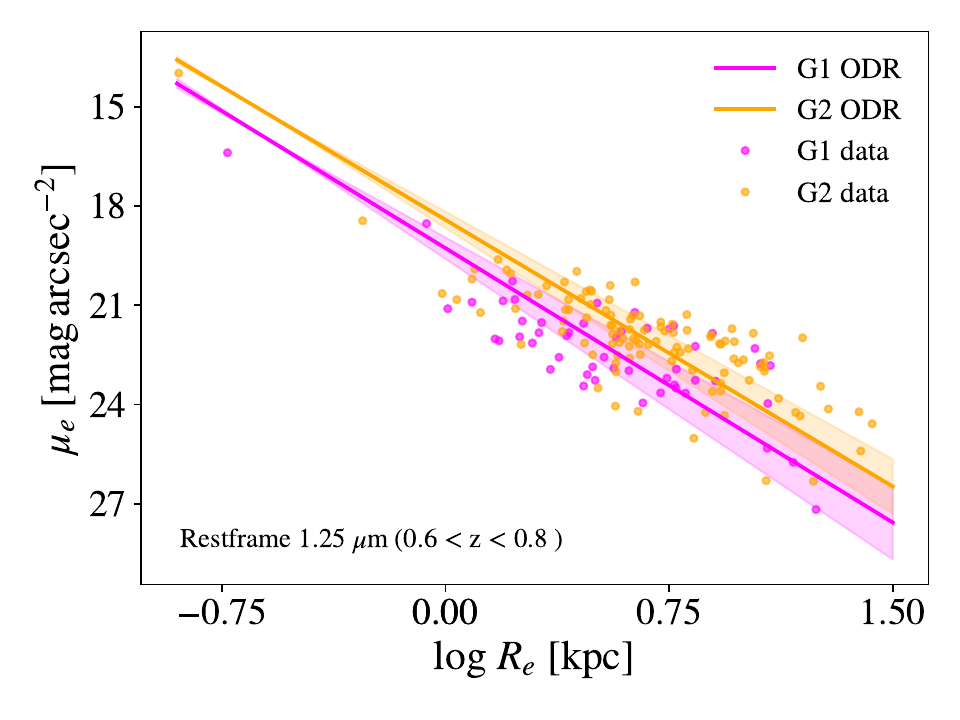}
    \caption{Examples of Kormendy relations fitted on galaxies in a particular redshift bin ($0.6 < z < 0.8$). The top panel shows the fit for the F814W filter, while the bottom panel shows the same for the rest-frame J-band. Solid lines show the fitted relations, while the data points are shown as circles. Shaded regions indicate uncertainties for the relations estimated using 1000 bootstrap rounds of fitting, adopting the formula $Q_\sigma = 0.74(Q75 - Q25)$, where Q25 and Q25 are the 25th and 75th percentiles of the fitting parameters.}
    \label{fig:KR_fit_example}
\end{figure}

In this appendix, we present additional figures related to the fitting of the Kormendy relation described in Section \ref{sec:KR_method} and presented in Section \ref{sec:struct_relations}. In Figure \ref{fig:KR_fit_example}, we present an example of the resulting lines from the fitting of the KR described in Section \ref{sec:KR_method}. Additionally, we also show the evolution of the slope corresponding to the intercepts of Figure \ref{fig:kormendy}.

\begin{figure}
    \centering
    \includegraphics[width=0.45\textwidth]{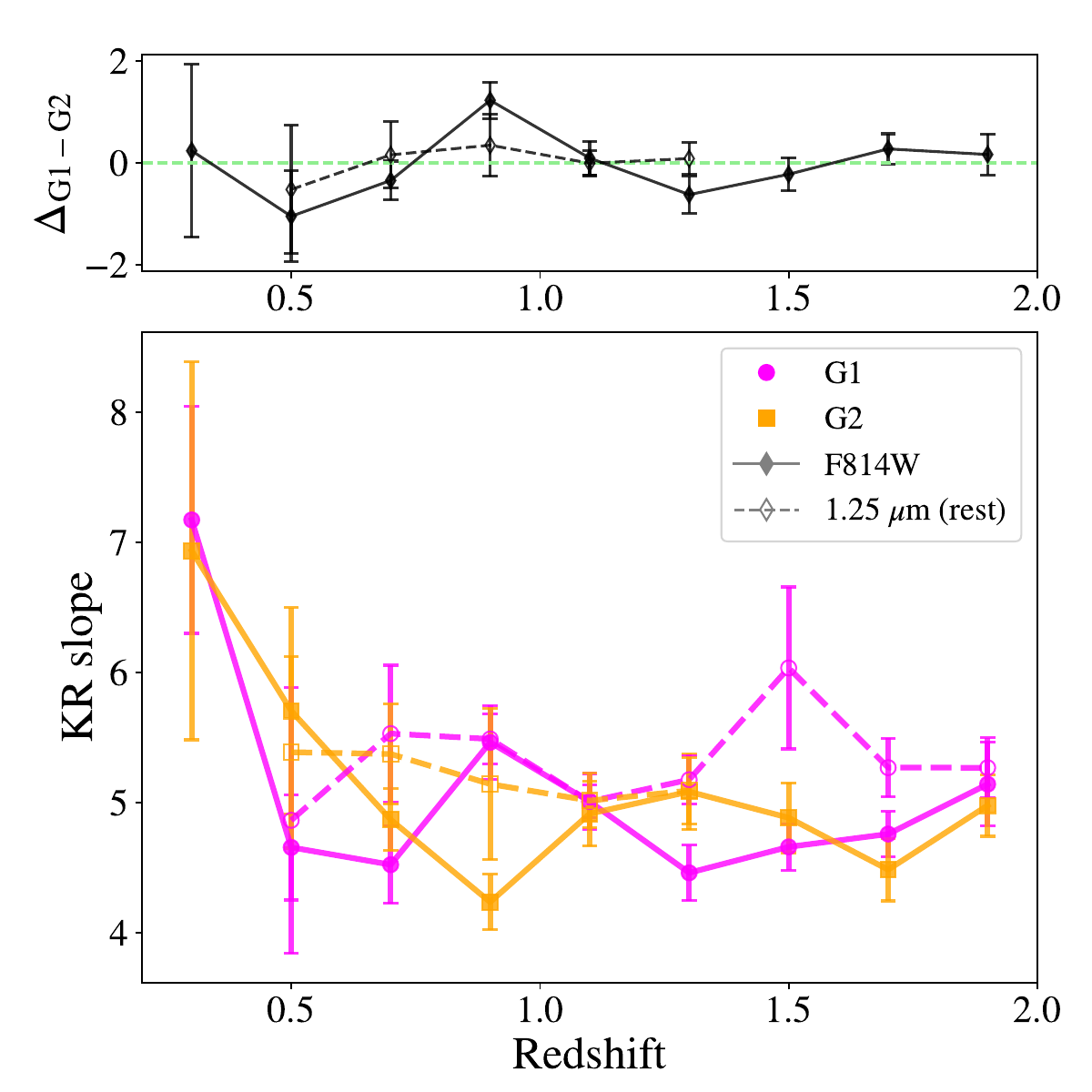}
    \caption{Slope ($a$) of the Kormendy relations fitted for \emph{G1} (magenta) and \emph{G2} (orange) galaxies as a function of redshift. The lower panel shows the evolution of $a$ calculated from observed F814W (solid) and rest-frame J-band (dashed) magnitudes, with uncertainties estimated using 1000 bootstrap rounds of fitting, adopting the formula $Q_\sigma = 0.74(Q75 - Q25)$, where Q25 and Q25 are the 25th and 75th percentiles of $a$. The top panel shows the difference between \emph{G1} and \emph{G2} (black lines) in each redshift bin. In the top panel, the line styles of the black lines follow the line styles in the bottom panel, and the green dashed horizontal line indicates no difference between the intercepts. Error bars are estimated by propagation of uncertainty, addition in quadrature in this case.}
    \label{fig:KR_slope}
\end{figure}

\section{Morphological metrics and Sérsic parameters across redshift}
\label{app:morph}

\begin{figure*}
    \centering
    \includegraphics[width=\linewidth]{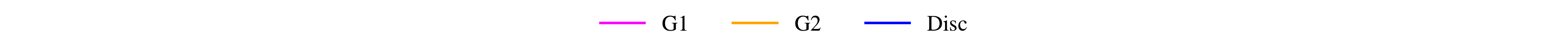}
    \includegraphics[width=\linewidth]{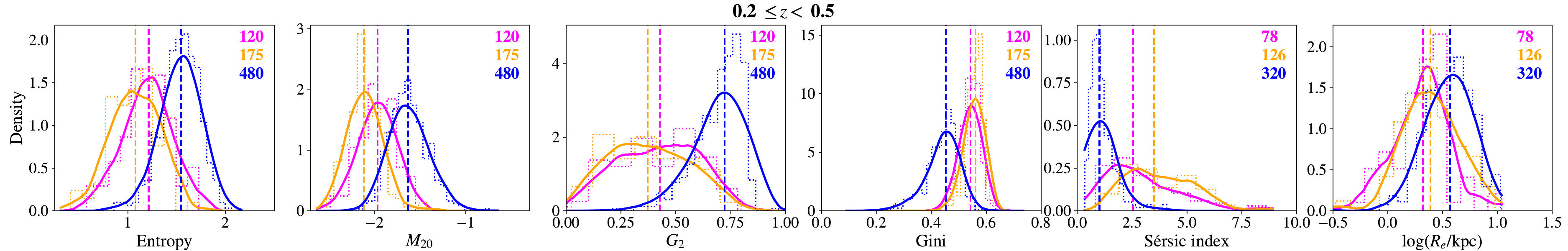}
    \includegraphics[width=\linewidth]{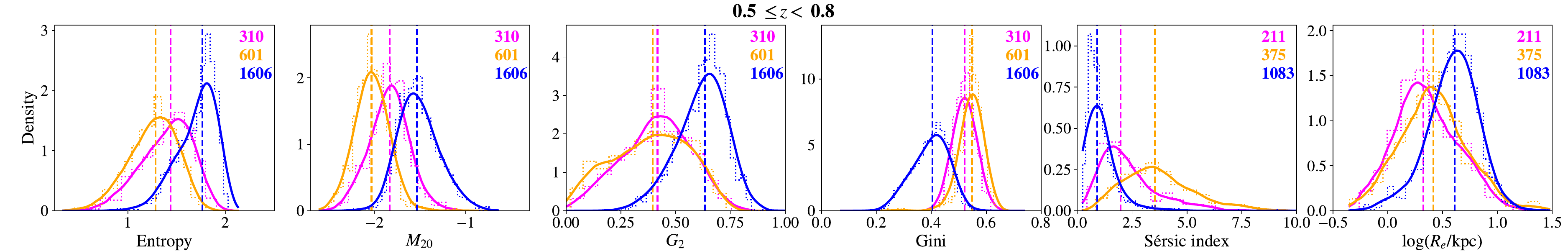}
    \includegraphics[width=\linewidth]{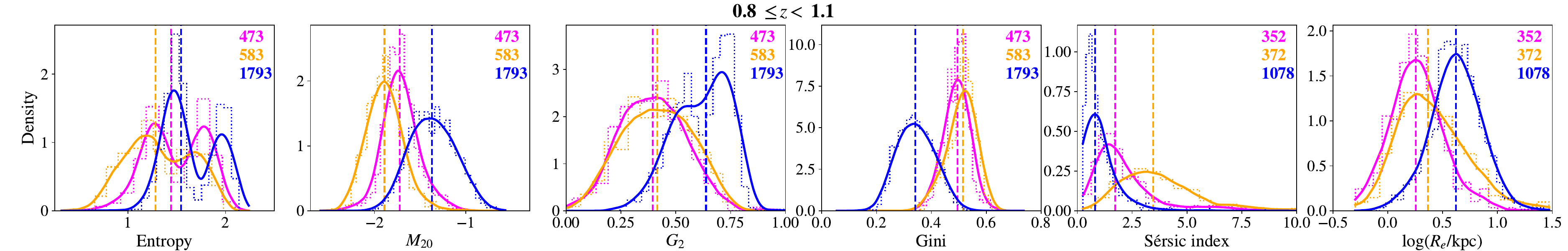}
    \includegraphics[width=\linewidth]{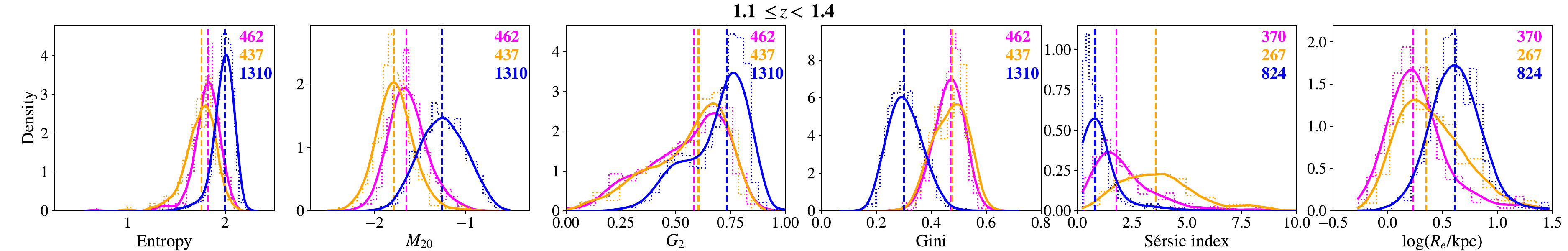}
    \includegraphics[width=\linewidth]{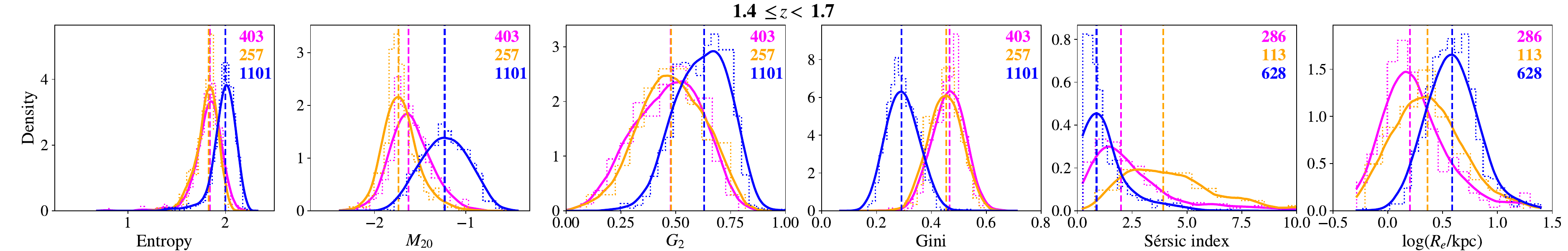}
    \includegraphics[width=\linewidth]{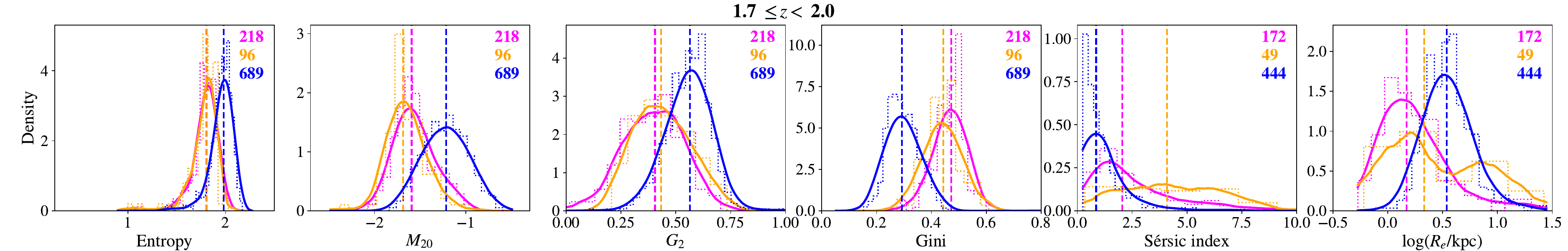}
    \includegraphics[width=\linewidth]{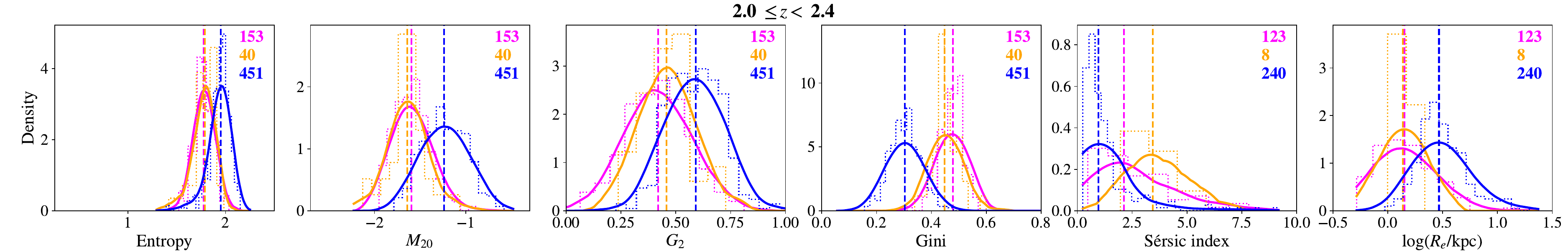}
    \caption{Distribution of morphological metrics and parameters in different redshift bins. Colours are the same as in Figure \ref{fig:morph_metrics}, with the numbers in the top right indicating the number of galaxies in each class at that redshift. Solid lines indicate Epanechnikov kernel density estimations (KDEs) and dotted lines indicate histograms with bins defined by the Freedman-Diaconis rule. The vertical axes show probability density, so the area under the curves or histograms sums to 1. Vertical dashed lines indicate medians for each sample.} 
    \label{fig:morph_across_z}
\end{figure*}

In Section \ref{sec:morph_metrics}, we show that \emph{G1} galaxies are intermediate to \emph{G2} and disc galaxies in almost all metrics of the MEGG system (see Figure \ref{fig:morph_metrics}). To test the influence of redshift on the results that we present in Section \ref{sec:morph_metrics}, we now divide the \emph{G1}, \emph{G2} and disc galaxies into redshift bins, and present the distribution of their MEGG metrics in Figure \ref{fig:morph_across_z}. We also present the parameter distributions from Sérsic fits as additional information to reinforce our results.
As shown in most panels (especially at $z < 1.4$), the general trend of \emph{G1} being intermediate persists. Additionally, we observe that \emph{G1} galaxies lie between discs and \emph{G2} galaxies in terms of the Sérsic index. 

\end{document}